\documentclass{aa}  
\usepackage[varg]{txfonts}

\usepackage{bm} 

\usepackage[T1]{fontenc}
\usepackage{ae,aecompl}
\usepackage{graphicx}   
\usepackage{amsmath}    
\usepackage{amssymb}    

\usepackage[x11names]{xcolor}

\usepackage{amsmath}	
\usepackage{mathtools}
\usepackage{amssymb}	
\usepackage{natbib}
\bibpunct{(}{)}{;}{a}{}{,} 

\usepackage[draft]{hyperref}

\hypersetup{
   colorlinks=true,
   linkcolor=blue,
   citecolor=blue,
   urlcolor=blue,
}

\usepackage{enumerate}

\newcommand{\mstar}{$M_{\rm star}$}
\newcommand{\mbreak}{$M_{\rm break}$}
\newcommand{\mgas}{$M_{\rm gas}$}
\newcommand{\logoh}{12$+$log(O/H)}
\newcommand{\lco}{$L_{\rm CO}^\prime$}
\newcommand{\mhi}{$M_{\rm HI}$}
\newcommand{\mhtwo}{$M_{\rm H2}$}
\newcommand{\aco}{$\alpha_{\rm CO}$}
\newcommand{\coone}{$^{12}$CO($1-0$)}

\newcommand{\kkmspc}{$({\rm K\,km\,s^{-1}\,pc^{2}})^{-1}$}
\newcommand{\lcounits}{${\rm K\,km\,s^{-1}\,pc^{2}}$}

\newcommand{\pcatwo}{3DPCA$^2$(OH)}
\newcommand{\pcaone}{3DPCA$^1$--OH}

\newcommand{\micron}{$\mu$m}

\newcommand{\zsun}{$Z_\odot$}
\newcommand{\msun}{M$_\odot$}

\newcommand{\ergs}{erg\,s$^{-1}$}
\newcommand{\msunyr}{M$_\odot$\,yr$^{-1}$}
\newcommand{\htwo}{H$_2$}
\newcommand{\hi}{H{\sc i}}
\newcommand{\hidef}{H{\sc i}--def}
\newcommand{\hii}{H{\sc ii}}

\newcommand{\av}{$A_{\rm V}$}

\newcommand{\ha}{H$\alpha$}
\newcommand{\hb}{H$\beta$}

\newcommand{\oiii}{[O{\sc iii}]}

\newcommand{\sii}{[S{\sc ii}]}
\newcommand{\nii}{[N{\sc ii}]}
\newcommand{\te}{T$_{\rm e}$}

\newcommand{\hers}{{\it Herschel}}

\begin{document}

\title{Scaling~relations~and~baryonic~cycling~in~local~star-forming~galaxies: I. The sample}
\titlerunning{Scaling relations and baryonic cycling in local star-forming galaxies: The sample}

\author{M. Ginolfi\inst{\ref{inst1}}
	\and
	L. K. Hunt\inst{\ref{inst2}}
	\and
	C. Tortora\inst{\ref{inst2}}
	\and
	R. Schneider\inst{\ref{inst3}}
	\and
	G. Cresci\inst{\ref{inst2}}
}

\institute{Observatoire de Gen\`eve, Universit\`e de Gen\`eve, 51 Ch. des Maillettes, 1290 Versoix, Switzerland\\
	\email{michele.ginolfi@unige.ch}\label{inst1}
	\and
	INAF/Osservatorio Astrofisico di Arcetri, Largo Enrico Fermi 5, I-50125 Firenze, Italy\label{inst2}
	\and
	Dipartimento di Fisica, Sapienza Universit\`a di Roma, Piazzale Aldo Moro 5, I-00185, Roma, Italy\label{inst3}
}

\date{Received XXX; accepted YYY}

\abstract{Metallicity and gas content are intimately related in the baryonic exchange cycle of galaxies, and galaxy evolution scenarios
can be constrained by quantifying this relation.
To this end, we have compiled a sample of $\sim$400 galaxies in the Local Universe, dubbed ``MAGMA'' (Metallicity And Gas for Mass Assembly),
which covers an unprecedented range in parameter space,
spanning more than 5 orders of magnitude in stellar mass (\mstar), star-formation rate (SFR), and gas mass (\mgas), 
and a factor of $\sim 60$ in metallicity [$Z$, \logoh].
Stellar masses and SFRs have been recalculated for all the galaxies using IRAC,
WISE and GALEX photometry, and \logoh\ has been transformed, where necessary, to a common metallicity calibration. 
To assess the true dimensionality of the data, we have applied multi-dimensional principal component analyses (PCAs) to our sample.
In confirmation of previous work, we find that even with the vast parameter space covered by MAGMA, the relations
between \mstar, SFR, $Z$ and \mgas\ (\mhi$+$\mhtwo) require only two dimensions to describe the hypersurface.
To accommodate the curvature in the \mstar--$Z$ relation, we have applied a piecewise 3D PCA
that successfully predicts observed \logoh\ to an accuracy of $\sim 0.1$\,dex.
MAGMA is a representative sample of isolated star-forming galaxies in the Local Universe,
and can be used as a benchmark for cosmological simulations and to calibrate evolutionary trends with redshift.

}


\maketitle

\section{Introduction}
\label{sec:introduction}

As long as star formation occurs in their gas reservoirs, galaxies evolve increasing their stellar mass (\mstar) and their metal content, 
depending on the relative efficiency of inflows/outflows, dynamical interactions, and environmental processes. 
In other words, at any time, \mstar\ and metallicity ($Z$) reflect the combined effect of both the integrated history of star formation and the degree of interaction with the surrounding environment.
Not surprisingly, the causal links between gas mass (\mgas), star formation rate (SFR), \mstar, and $Z$, manifest in a number of observed correlations between these quantities, 
often referred to as {\it scaling relations}. 
Some among the most notable examples are: 
{\it (i)} the correlation between \mstar\ and SFR 
\citep[dubbed the ``Main Sequence'', MS: e.g.,][]{Brinchmann2004, Noeske2007a, Daddi2010a, Elbaz2011, Renzini2015};
{\it (ii)}
the correlation between \mgas\ and SFR 
\citep[the ``Schmidt-Kennicutt", SK, relation; e.g.,][]{Schmidt1959, Kennicutt1998, Bigiel2008, Leroy2009};
and {\it (iii)}
the ``mass-metallicity relation", MZR, between \mstar\ and $Z$ 
\citep[e.g.,][]{Lequeux1979, Tremonti2004, Maiolino2008}.
In star-forming galaxies,
$Z$ is typically measured by the abundance of oxygen, O/H, in the ionized gas, as it is the most abundant heavy element produced by massive stars.

These scaling relations among fundamental properties of galaxies are potentially insightful tools to explore demographics of galaxies and their evolution. 
In particular, the mutual correlations among physical properties in galaxies imply that the observed residuals from the main relations 
(in other words, their intrinsic scatters) could be correlated with other variables.
Many studies have investigated such a notion, and this type of analysis has proved to be a powerful diagnostic, 
providing simple quantitative tests for analytical models and numerical simulations.

Only recently has it been possible to incorporate gas properties in studies of baryonic cycling, 
thanks to the growing number of available gas measurements (atomic and molecular), including:
the Arecibo Legacy Fast ALFA Survey \citep[ALFALFA,][]{Haynes2011,Haynes2018}
the Galaxy Evolution Explorer (GALEX) Arecibo SDSS Survey \citep[GASS,][]{Catinella2010,Catinella2018};
the COLD-GASS survey \citep{Saintonge2011a,Saintonge2017};
the Nearby Field Galaxy Survey \citep[NFGS,][]{Jansen2001,Wei2010,Stark2013};
the \hers\ Reference Survey \citep[HRS,][]{Boselli2010,Cortese2011,Boselli2014a};
and the APEX Low-redshift Legacy Survey for MOlecular Gas \citep[ALLSMOG,][]{Bothwell2014,Cicone2017}.
These surveys have provided important new observations of \hi\ and CO, in order to derive \htwo\ and compare
gas content with other galaxy properties.
Results suggest that the relation of atomic gas to \mstar\ and SFR drives 
a galaxy's position relative to the MS \citep[e.g.,][]{Huang2012,Gavazzi2013,Saintonge2016},
and that \hi\ gas fractions increase with decreasing \mstar\ and stellar mass surface density, 
at least down to log(\mstar/\msun)\,=\,9 \citep[e.g.,][]{Cortese2011,Gavazzi2013,Catinella2018}.
Incorporating molecular gas \htwo\ in the analysis suggests that the strongest correlations are
between \htwo\ content and SFR;
in particular, molecular depletion time depends strongly on specific SFR (sSFR\,$\equiv$\,SFR/\mstar)
\citep[e.g.,][]{Saintonge2011a,Saintonge2011b,Boselli2014b,Hunt2015,Saintonge2017}.

Important clues to baryonic cycling also come from systematic studies of the intrinsic scatter of the MZR, finding that a 
{\it fundamental metallicity relation}
(FMR) exists between \mstar, $Z$ and SFR, that minimizes the scatter in the MZR 
\citep[see e.g.,][]{Ellison2008, Mannucci2010}. 
According to the FMR, galaxies lie on a tight, redshift-independent two-dimensional (2D) surface in 
3D space defined by \mstar, $Z$ and SFR, where at a given \mstar, galaxies with higher SFR have systematically lower gas-phase $Z$ 
\citep[see e.g.,][]{Hunt2012,LaraLopez2013,Hunt2016a,Hashimoto2018,Cresci2018}.
Many theoretical models have  investigated this finding, explaining it in terms of an equilibrium between metal-poor inflows and metal-enriched outflows 
\citep[e.g.,][]{Dave2012,Dayal2013,Lilly2013,Graziani2017}.
Observational results suggest that the FMR may be more strongly expressed via the gas mass rather than via the SFR \citep[see e.g.,][]{Bothwell2013,Brown2018}. 
In this light, the FMR might be interpreted as a by-product of an underlying relationship between the scatter of the MZR and the gas content 
\citep[e.g.,][]{Zahid2014}. 
In particular, 
\citet{Bothwell2016b}, with an analysis that included \mstar, SFR, O/H, and molecular gas mass, \mhtwo, 
suggest that the \textit{true} 
FMR exists between \mstar, O/H and \mhtwo, which is linked to SFR via the SK star-formation law.

Virtually all previous studies of gas scaling relations in galaxies have focused on galaxies
more massive than $10^{9}$\,\msun. 
In this paper, we extend previous studies to lower stellar masses, 
reporting the analysis of the mutual dependencies of physical properties in a sample 
of $\sim 400$ local galaxies, 
with simultaneous availability of \mstar, SFR, \mhi, \mhtwo\ (thus also total gas, \mgas),  
and O/H, spanning an unprecedented range in \mstar, 
from $\sim 10^{5}$\,\msun\ to $3\times10^{11}$\,\msun.
In Sect. \ref{sec:sample}, we first describe the individual sub-samples, and
then homogenize the stellar mass and SFR estimates by incorporating
mid-IR (MIR) fluxes from the Wide-field Infrared Survey Explorer \citep[WISE,][]{Wright2010}
and photometry from GALEX \citep{Morrissey2008}.
With updated principle component analysis (PCA) techniques, 
Sect. \ref{sec:pca} explores the correlations in the 
four- and three-dimensional (4D, 3D)
parameter spaces defined by \mstar, SFR, O/H, and \mgas, 
together with the two separate gas components \mhi\ and \mhtwo.
There is a particular focus in Sect. \ref{sec:comparison}
on the MZR scatter and the ramifications of including a significant population of low-mass galaxies in the sample.

\section{Combined sample: MAGMA}
\label{sec:sample}

We have compiled a sample of 392 local galaxies, with simultaneous availability of \mstar, SFR,  gas masses 
(both atomic, \mhi, and molecular, \mhtwo, the latter obtained by measurements of CO luminosity, \lco) and metallicities [\logoh].
We assembled our sample by combining a variety of previous surveys at $z \sim 0$ with new observations of CO in low-mass galaxies.
The details of the parent surveys such as metallicity calibration, stellar-mass, and SFR determinations are provided below.
The following four selection criteria are adopted:
\begin{enumerate}[(1)]
\item 
{\it only} galaxies with robust ($\gtrsim 3\sigma$) 
detections of \mstar, SFR, \logoh, \mhi, and \lco\ are considered; 
\item 
galaxies were eliminated if they were thought to host active galactic nuclei (AGN) based on the 
BPT classifications\footnote{The Baldwin-Philips-Terlevich (BPT) diagram classification \citep{Baldwin1981} relies on the emission-line properties of galaxies, based on the 
\sii/\ha\ versus \oiii/\hb\ ratios.} provided by the original surveys;
\item
when \hi-deficiency measurements \hidef\ were available \citep[e.g.,][]{Boselli2009,Boselli2014a}, 
following \citet{Boselli2014b}, 
only galaxies  with \hidef\,$\leq$0.4 were retained; 
\item 
properties of galaxies in common among two or more parent surveys have been taken from the sample that provided more ancillary information 
(e.g., high quality spectra, resolved maps, uniform derivation of parameters, etc.).
\end{enumerate}

\hi-deficiency is defined as the logarithm of the ratio of the observed \hi\ mass of a galaxy
and the mean \hi\ mass expected for an isolated galaxy with the same optical size and morphological type \citep[e.g.,][]{Haynes1984}.
The \hi-deficiency requirement is included to ensure that our sample is representative of isolated, field galaxies,
not having undergone potential stripping effects from residence in a cluster.
Because we require metallicity and gas measurements, we have dubbed our compiled sample 
MAGMA (Metallicity And Gas in Mass Assembly).
The final MAGMA sample has been drawn from the following nine parent surveys/papers:

\begin{itemize}
\renewcommand\labelitemi{--}	
\setlength\itemsep{0.5\baselineskip}
\item 
	{\bf xGASS-CO: } xGASS-CO is the overlap between 
	the extended GALEX Arecibo SDSS Survey \citep[xGASS:][]{Catinella2018} and the extended CO Legacy Database for GASS 
	\citep[xCOLD GASS:][]{Saintonge2017}.
	xGASS\footnote{The full xGASS representative sample is available on the xGASS website, 
\url{http://xgass.icrar.org} in digital format.}
is a gas fraction-limited census of the \hi\ gas content of $\sim1200$ local galaxies, spanning over 2 decades in stellar mass 
(\mstar$\,=\,10^{9}-10^{11.5}$\,\msun). 
	The xCOLD GASS survey\footnote{The full xCOLD GASS survey data products are available on the xCOLD GASS website 
\url{http://www.star.ucl.ac.uk/xCOLDGASS/}.} 
contains IRAM-30m CO(1-0) measurements for 532 galaxies also spanning the entire SFR-\mstar\ plane at \mstar$> 10^9$\,\msun.
	Stellar masses are from the MPA-JHU\footnote{\href{http://wwwmpa.mpa-garching.mpg.de/SDSS/DR7/}{http://wwwmpa.mpa-garching.mpg.de/SDSS/DR7/}} 
	catalogue, where \mstar\ is computed from a fit to the 
spectral energy distribution (SED) obtained using SDSS broad-band photometry \citep{Brinchmann2004, Salim2007}.
	SFRs are computed as described by \citet{Janowiecki2017} by combining NUV with mid-IR (MIR) fluxes from the Wide-field Infrared Survey Explorer 
\citep[WISE;][]{Wright2010}.
When these are not available (the case for $\sim$70\% of the xGASS sample), SFRs are determined using a ``ladder'' technique \citep{Janowiecki2017,Saintonge2017}.
Data Release 7 \citep[SDSS DR7,][]{Abazajian2009}, 
calibrated by \citet{Saintonge2017} to the \nii-based strong-line calibration by \citet[][PP04N2]{Pettini2004}.
In addition to omitting AGN and Seyferts (see above), we have also excluded galaxies in \citet{Saintonge2017} with an ``undetermined" 
or ``composite" classification, based on the BPT diagram; metallicities from PP04N2 for such galaxies tend to be highly uncertain.
xGASS-CO, the overlap  between xGASS and xCOLDGASS, includes 477 galaxies, with 221 non-AGN galaxies with robust CO detections. 
The subset of xGASS-CO that respects our selection criteria (i.e., with \hi\ and CO detections and not excluded for potentially uncertain O/H calibration) 
consists of {\bf 181} galaxies.
\item
	{\bf HRS: } The \hers\ 
Reference Sample \citep{Boselli2010} is a $K$-band selected, volume-limited sample comprising 323 galaxies. 
HRS\footnote{A full description of the survey and the ancillary data can be found at \url{https://hedam.lam.fr/HRS/.}} is a fairly complete
description of the Local Universe galaxy population although underrepresented in low-mass galaxies \citep[see][]{Boselli2010}.
Stellar masses \mstar\ and SFRs were obtained from \citet{Boselli2015}, where
\mstar\ values were derived according to the precepts of \citet{Zibetti2009} using $i$-band luminosities and $g-i$ colors, and
SFRs are the mean of the four methods investigated by \citet{Boselli2015}.
These include radio continuum at 20\,cm, \ha$+$24\,\micron\ luminosities, FUV$+$24\,\micron\ luminosities, and \ha\ luminosities corrected for
extinction using the Balmer decrement\footnote{\citet{Boselli2015} do not publish the individual estimates, so we were unable to
select the hybrid method based on 24\,\micron\ luminosities that would be more consistent with other samples discussed  here.}.
\logoh\ was taken from \citet{Hughes2013} based on the PP04N2 calibration,
and gas quantities, \mhi\ and \lco\ were taken from \citet{Boselli2014a}.
86 HRS galaxies have \hi\ and CO detections, but 18 of these have \hidef$>$0.4 \citep[as given by][]{Boselli2014a}, so 
we are left with {\bf 68} HRS galaxies that satisfy our selection criteria.
\item 
	{\bf ALLSMOG: } The APEX Low-redshift Legacy Survey of MOlecular Gas \citep[][]{Bothwell2014,Cicone2017}
		comprises 88 nearby, star-forming galaxies with stellar masses in the range $10^{8.5} < M_{\rm star}/M_\odot < 10^{10}$, and gas-phase metallicities \logoh\,$>$\,8.4. 
	ALLSMOG\footnote{The full ALLSMOG survey data products are available on the web page \url{http://www.mrao.cam.ac.uk/ALLSMOG/.}}
is entirely drawn from the MPA-JHU catalogue of spectral measurements and galactic parameters of SDSS DR7.
	Stellar mass and SFR values of ALLSMOG galaxies are taken from the MPA-JHU catalogue, and the SFR is based on the 
(aperture- and extinction-corrected) \ha\ intrinsic line luminosity. 
We have used the PP04N2 O/H calibration given by \citet{Cicone2017}.
To convert the ALLSMOG CO(2--1) values from \citet{Cicone2017} to the lower-J CO(1--0) available for the remaining samples, 
we assume $R_{21}\,=\,0.8$ as they advocate. 
	The subset of ALLSMOG that respects our selection criteria consists of {\bf 38} galaxies.
\item 
	{\bf KINGFISH: } The Key Insights on Nearby Galaxies: a Far-Infrared Survey with \hers, 
KINGFISH\footnote{An overview of the scientific strategy for KINGFISH and the properties of the galaxy sample can be found on the web page 
\url{https://www.ast.cam.ac.uk/research/kingfish}.} \citep{Kennicutt2011},
	contains 61 galaxies with metallicity in the range $7.54\,\leq$\,\logoh\,$\leq\,8.77$ and stellar masses in the range $[2\times10^7 - 1.4\times10^{11}]\,$\msun.
	Stellar masses and SFRs are taken from \citet{Hunt2019}.
The \mstar\ values were computed from the SFR-corrected IRAC 3.6\,\micron\ luminosities according to 
the luminosity-dependent mass-to-light (M/L) ratio given by \citet{Wen2013}, and
are within $\sim$0.1\,dex of those derived by comprehensive SED fitting \citep[see][]{Hunt2019}.
SFRs are inferred from the far-ultraviolet (FUV) luminosity combined with total-infrared (TIR) luminosity following \citet{Murphy2011}. 
Atomic gas masses \mhi\ and CO measurements for \mhtwo\ are taken from \citet{Kennicutt2011}, with refinements from \citet{Sandstrom2013} and \citet{Aniano2020}.
``Representative'' metallicities evaluated at 0.4 times the optical radius R$_{\rm opt}$ 
from \citet{Moustakas2010} were converted from the \citet[][KK04]{Kobulnicky2004} to the PP04N2 calibration according
to the transformations given by \citet{Kewley2008}; more details are given in \citet{Hunt2016a} and \citet{Aniano2020}. 
After omitting NGC\,2841 and NGC\,5055, because their metallicities exceeded the valid regime for the \citet{Kewley2008} transformations,
the required data are available for
38 KINGFISH galaxies. 
Three of these have \hidef$>$0.4 \citep[given by][]{Boselli2014a} so we ultimately select {\bf 35} galaxies from KINGFISH.
\item 
	{\bf NFGS: } The Nearby Field Galaxy Survey \citep[][]{Jansen2000,Kewley2005,Kannappan2009} consists of 196 galaxies
spanning the entire Hubble sequence in morphological types, and a range in luminosities from low-mass dwarf galaxies to luminous massive systems.
Stellar masses, \mstar, are given by \citet{Kannappan2013} and are based on NUV$+${\it ugrizJHK}$+$IRAC 3.6\,\micron\ SEDs.
We have taken (spatially) integrated SFR (based on \ha) and O/H values from \citet{Kewley2005}, and transformed \logoh\ from their \citet[][KD02]{Kewley2002} calibration to PP04N2 
according to the formulations by \citet{Kewley2008}.
\citet{Stark2013} provide CO and \mhtwo\ measurements, and \mhi\ is tabulated by \citet{Wei2010} and \citet{Kannappan2013}.
After removing NGC\,7077, that appears in the following dwarf sample, there are {\bf 26} galaxies
that meet our selection criteria. 
\item 
	{\bf BCDs: } The Blue Compact Dwarf galaxies (BCDs) have been observed and detected in \coone\ with the IRAM 30m single dish 
\citep{Hunt2015,Hunt2017}. 
They were selected primarily from the primordial helium sample of \citet{Izotov2007}, known to have reliable metallicities \logoh\
measured through the direct electron-temperature (\te) method.
An additional, similar, set of BCDs has been detected in \coone\ (Hunt et al. 2020, in prep.) with analogous selection criteria.
For both sets of BCDs,
\mstar\ is derived as for KINGFISH galaxies, namely from IRAC 3.6\,\micron\ or WISE 3.4\,\micron\ luminosities, after correcting for
free-free, line emission based on SFR, and dust continuum when possible \citep[see also][]{Hunt2015}.
This method has been shown to be consistent with full-SED derived \mstar\ values
to within $\la$0.1\,dex \citep{Hunt2019}.
For the galaxy in common with the NFGS, NGC\,7077, the two \mstar\ estimates are the same to within 0.07\,dex.
SFRs are based on the \citet{Calzetti2010} combination of \ha\ and 24\,\micron\  luminosities.
\hi\ masses are given by 
\citet{Hunt2015} and Hunt et al. (2020, in prep.).
As mentioned above, \logoh\ is obtained from the direct \te\ method \citep[for details see][]{Hunt2016a}.
The subset of BCDs that respect our selection criteria with \hi\ and CO [\coone] detections comprises 
{\bf 17} galaxies with metallicities \logoh\ ranging from 7.7 to 8.4;
to our knowledge, this is the largest sample of low-metallicity dwarf galaxies in the Local Universe detected in CO. 
	\item 
	{\bf DGS: } The Dwarf Galaxy Survey, 
DGS\footnote{Information on the DGS sample, as well as data products, can be found on the website 
\url{http://irfu.cea.fr/Pisp/diane.cormier/dgsweb/publi.html}.} \citep{Madden2013,Madden2014},
is a \hers\ 
sample of 48 local metal-poor low-mass galaxies, 
with metallicities ranging from \logoh\ = 7.14 to 8.43 and stellar masses from $3\times10^6$ to $\sim 3 \times 10^{10}~M_\odot$.
The DGS sample was originally selected from several deep optical emission line and photometric surveys 
including the Hamburg/SAO Survey and the First and Second Byurakan Surveys \citep[e.g.,][]{Ugryumov1999,Ugryumov2003,Markarian1983,Izotov1991}.
Although stellar masses are given by \citet{Madden2013} with corrected values in \citet{Madden2014},
these are calculated according to \cite{Eskew2012} using the Spitzer/IRAC luminosities at 3.6\,\micron\ and 4.5\,\micron.
\citet{Hunt2016a} gives \mstar\ for these same galaxies by first subtracting nebular continuum and line emission, known to be important in metal-poor star-forming
dwarf galaxies \citep[e.g.,][]{Smith2009};
a comparison shows that the \mstar\ values in \citet{Madden2014} are, on average, 0.3\,dex larger than those by \citet{Hunt2016a}.
Thus, in order to maximize consistency with the other samples considered here,
like for the KINGFISH, BCDs, and the Virgo star-forming dwarfs (see below), we have used stellar masses based on WISE and/or IRAC 3.4-3.6\,\micron\
luminosities using the recipe by \citet{Wen2013} after subtracting off non-stellar emission estimated from the SFR
\citep[see also][]{Hunt2012,Hunt2015,Hunt2019}.
Metallicities for the DGS are taken from \citet{DeVis2017}, using their PP04N2 calibration.
We have also recalculated the SFRs using \ha\ and 24\,\micron\ luminosities as advocated by \citet{Calzetti2010} and reported in \citet{Hunt2016a}.
Of the 48 DGS galaxies discussed by \citet{RemyRuyer2014}, 7 are included in the BCD sample observed in CO by \citet{Hunt2015,Hunt2017}; 
5 have CO detections from \citet{Cormier2014}; and 9 elsewhere in the literature
\citep{Kobulnicky1995,Young1995,Greve1996,Walter2001,Leroy2005,Leroy2006,Gratier2010,Schruba2012,Oey2017}.
However, one of these, UM\,311, is a metal-poor \hii\ region within a larger galaxy \citep[see][]{Hunt2010}.
There is a discrepancy between the \mstar\ values given by \citet{Madden2014} and \citet{Hunt2016a} of
more than a factor of 100; this is roughly the difference between the larger UM\,311 complex and the individual
metal-poor \hii\ regions, illustrating that the gas content of the individual \hii\ regions is highly uncertain.
We thus eliminate UM\,311 from the DGS subset, and include the remaining {\bf 13} DGS galaxies
that respect our selection criteria. 
\item 
	{\bf Virgo star-forming dwarfs: } This sample of star-forming dwarf galaxies (SFDs) 
in Virgo is taken from a larger survey, namely the 
\hers\ Virgo Cluster Survey, 
HeViCS\footnote{Information on HeViCS and public data can be found in \url{http://wiki.arcetri.astro.it/HeViCS/WebHome}.} \citep[][]{Davies2010}.
Here to supplement the low-mass portion of our combined sample, we have included the dwarf galaxies in HeViCS studied by
\citet{Grossi2015, Grossi2016}. 
They were selected from the larger sample by requiring a dwarf morphology (e.g., Sm, Im, BCD) and detectable far-infrared (FIR)
emission with \hers. 
\mstar\ was estimated according to the WISE 3.4\,\micron\ luminosities,
and SFRs were calculated using \ha\ luminosities and correcting for dust using WISE 22\,\micron\ emission
as proposed by \citet{Wen2014}.
Both quantities are originally given with a \citet{Kroupa2001} initial mass function (IMF), and have been corrected
here to a \citet{Chabrier2003} IMF according to \citet{Speagle2014}.
Metallicities, \logoh, were based on the SDSS spectroscopy and use the PP04N2 calibration reported by \citet{Grossi2016},
with the exception of VCC\,1686 for which O/H was derived using the mass-metallicity relation given in \citet{Hughes2013}.
Of 20 targets observed, 11 were detected in \coone\ with the IRAM 30m \citep[][]{Grossi2016}.
Atomic hydrogen \hi\ is detected in all these \citep{Grossi2016}, 
but of the 11 galaxies with both \hi\ and CO detections, 4 have \hidef$>$0.4 \citep[from][]{Boselli2014a};
thus {\bf 7} Virgo SFDs satisfy our selection criteria.
\item 
	{\bf Extra single sources:} 
This subset includes individual galaxies that are not included in any survey, but for which our required
data of \mstar, SFR, O/H, CO, and \hi\ measurements exist.
These include 7 low-metallicity galaxies:
DDO\,53 and DDO\,70 (Sextans\,B) from \citet[][]{Shi2016}, 
NGC\,3310 \citep{Zhu2009},
NGC\,2537 \citep{GildePaz2002}, 
WLM \citep{Elmegreen2013},
Sextans\,A \citep{Shi2015}, 
NGC\,2403 \citep{Schruba2012}.
For these sources, as above for consistency, we used \mstar\ and SFR from \citet{Hunt2016a}.
We were able to compare global \mstar\ for one of these, WLM, and once reported to a common distance scale, our value of \mstar\ agrees with
that from \citet{Elmegreen2013} to within 0.1\,dex.
Metallicities for these objects are based on the direct \te\ method,
and taken from \citet{Engelbracht2008}, \citet{Marble2010}, and \citet{Berg2012}.
In the following figures, the {\bf 7} galaxies from these additional sources are labeled as ``Extra".
\end{itemize}

\begin{figure*}[!ht!]
\centering
\includegraphics[width=0.45\textwidth]{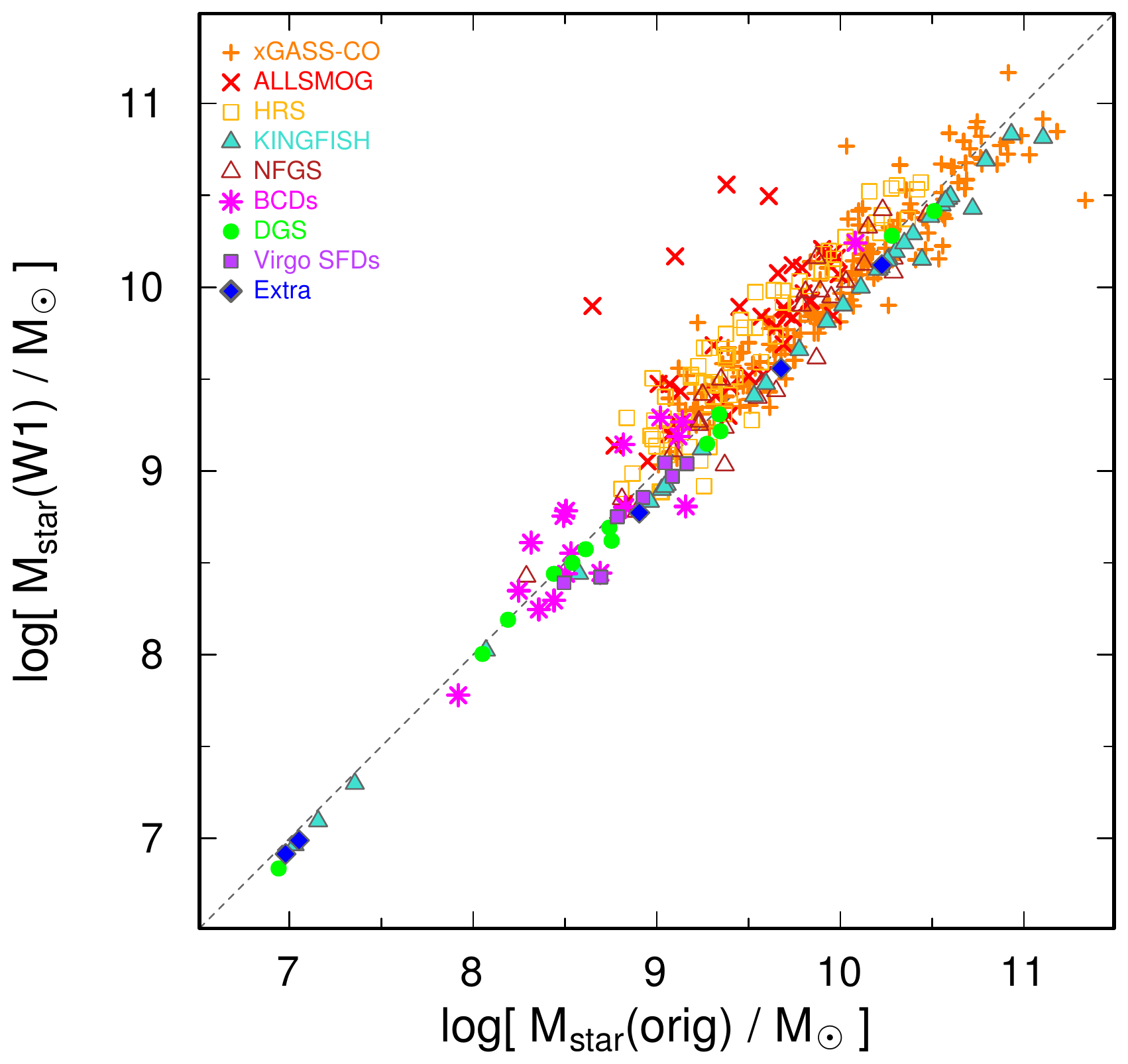}
\hspace{0.05\textwidth}
\includegraphics[width=0.45\textwidth]{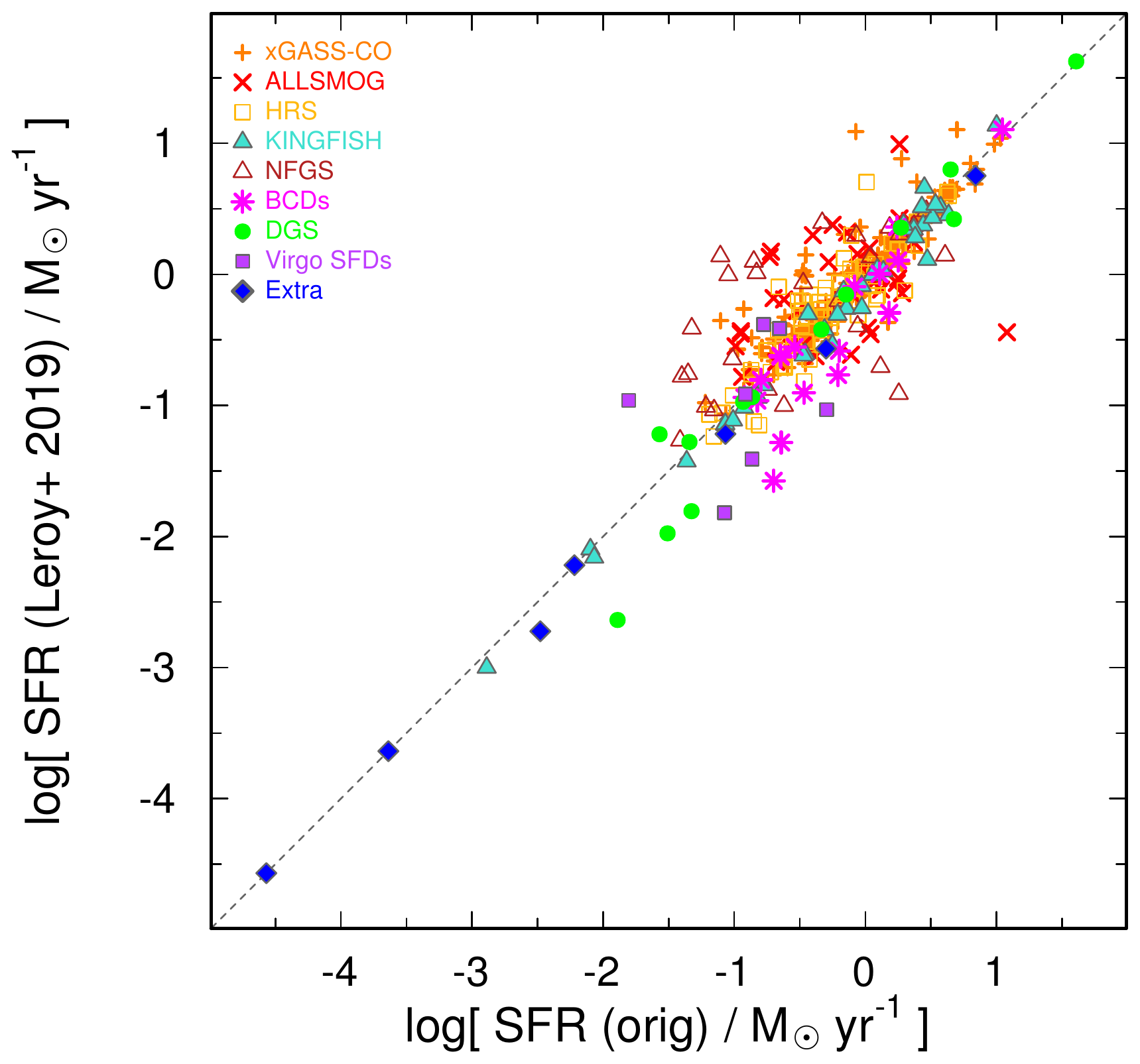}
\caption{Comparison of newly-derived \mstar\ (left panel) and SFR (right) with the original values.
The individual samples are distinguished by different symbols as given in the legend.
See text for more details on the derivations.
}
	\label{fig:compare}
\end{figure*}

\subsection{Galaxy parameters, data comparison, and potential selection effects}
\label{sec:detailed}

Because of potential systematics that could perturb our results,
we first ``homogenized'' the MAGMA sample by recalculating \mstar\ and SFR in a uniform way. 
The following sections compare the newly-derived values with the original ones described above.
We also analyze O/H, CO luminosities, and overall properties of the individual samples 
in order to assess any systematics that could affect the reliability of our results. 

\subsubsection{Stellar mass}

To estimate \mstar, 
we use $3.4-3.6$\,\micron\ luminosities, together with a luminosity-dependent M/L.
The photometry at 3.4\,\micron\ was acquired from the ALLWISE Source Catalogue
\citep[e.g.,][]{Wright2010},
taking the photometry measured in an elliptical aperture or within a circular aperture
of radius of 49\farcs5, whichever was larger.
Galactic extinction was corrected for using the
\citet{Schlafly2011} $A_B$ values\footnote{These were taken from the NASA/IPAC Extragalactic Database (NED),
funded by the National Aeronautics and Space Administration and operated by the California Institute of
Technology.} and the reddening curve by \citet{Draine2003}.
Luminosities were calculated from the apparent magnitudes according to the zero points of \citet{Jarrett2013}.
They were then converted to masses using the  M/L ratio for $3.4-3.6$\,\micron\ luminosities given by \citet{Hunt2019}, calibrated
on the CIGALE \citep{Boquien2019} stellar masses \citep[see also][]{Leroy2019}:
\begin{equation}
\log(M_{\rm star})\,=\,1.050\,\log(\nu L_\nu) + 0.387   .
\label{eqn:mstar}
\end{equation}
where $\nu L_\nu$ is the WISE (3.4\,\micron) or IRAC (3.6\,\micron) luminosities in units of \ergs.
Relative to the \mstar\ values obtained with detailed SED fitting,
Eqn. (\ref{eqn:mstar}) gives a slightly lower scatter and offset relative to the formulation of \citet{Wen2013},
and a negligible offset relative to constant M/L ratios at these wavelengths advocated
by various groups \citep[e.g.,][]{Eskew2012,Meidt2012,Meidt2014,Mcgaugh2014,Mcgaugh2015}.

When WISE photometry was unavailable, or had low signal-to-noise (mainly for the BCDs), we
used IRAC 3.6\,\micron\ photometry from \citet{Engelbracht2008} or from Hunt et al. (2020, in prep.).
We have assumed that IRAC 3.6\,\micron\ and WISE 3.4\,\micron\ monochromatic luminosities are identical to within uncertainties
as discussed in detail by \citet{Hunt2016a} and \citet{Hunt2019} \citep[see also][]{Leroy2019}.
All stellar masses are calculated according to a \citet{Chabrier2003} IMF \citep[for more details, see][]{Hunt2019}. 

Figure \ref{fig:compare} (left panel) compares the original \mstar\ values described in the preceding section
and the new ones derived here.
The mean differences (in log) are reported in Table \ref{tab:compare}.
There are apparently no systematics among the different samples, 
given that the deviations reported in Table \ref{tab:compare} are typically smaller or commensurate with their scatters.
However, there is some tendency for
the new WISE W1-derived \mstar\ to be larger than the originals, typically derived from optical SED fitting.
The best agreement is for xGASS-CO, but none of the samples, except for KINGFISH, shows a significant offset.
Moreover, for virtually all the samples the scatter is good to within a factor of 2.
This corresponds to an uncertainty of $\sim$0.3\,dex, consistent with the overall uncertainty of mass-to-light
ratios \citep[e.g.,][]{Hunt2019}.
The KINGFISH galaxies show the same offset relative to SED fitting results found by \citet{Hunt2019}.
Here we use the best-fit CIGALE-calibrated power-law slope with luminosity, and there the \citet{Wen2013}
power-law dependence was used; in any case, the scatter is small because the same photometry \citep[from][]{Dale2017} is adopted in both cases.

There are four ALLSMOG galaxies that show particularly high discrepancies
relative to our homogenized estimates of \mstar:
2MASXJ1336+1552, CGCG\,058$-$066, UGC\,02004, and VIII\,Zw\,039.
The previous stellar masses are roughly an order of magnitude smaller than the new values.
We have inspected the SDSS images of these, and they tend to be clumpy, with a series of
brightness knots throughout their disks. 
In these cases, the stellar masses automatically estimated by SDSS tend to regard the clumps, rather
than the galaxy as a whole.
If these galaxies are eliminated from the comparison of the new homogenized values and the
previous values for ALLSMOG, the mean difference (see Table \ref{tab:compare}) of old minus new log(\mstar) becomes 
$-0.18\,\pm\,0.16$.
These galaxies have been retained in our overall analysis.

\begin{table}[!t] 
\caption{Logarithmic differences of \mstar\ and SFR between the original description and adopted values$^{\mathrm a}$}
\label{tab:compare}
\resizebox{\linewidth}{!}{
\begin{tabular}{lrrr}
\hline
\\ 
\multicolumn{1}{c}{Sample} &
\multicolumn{1}{c}{log(\mstar/\msun)} &
\multicolumn{1}{c}{log(SFR/\msunyr)} &
\multicolumn{1}{c}{Number} \\
&
\multicolumn{1}{c}{[old $-$ new]} &
\multicolumn{1}{c}{[old $-$ new]} \\
\multicolumn{1}{c}{(1)} &
\multicolumn{1}{c}{(2)} &
\multicolumn{1}{c}{(3)} &
\multicolumn{1}{c}{(4)} \\
\\ 
\hline 
\\
xGASS-CO        & $-0.018\,\pm\,0.02$   &  $-0.045\,\pm\,0.19$ & 181  \\
ALLSMOG         & $-0.278\,\pm\,0.32$    &  $-0.083\,\pm\,0.46$ & 38  \\
HRS             & $-0.170\,\pm\,0.17$    &  $-0.024\,\pm\,0.20$ & 68 \\
KINGFISH        & $0.122\,\pm\,0.06$     &  $0.069\,\pm\,0.11$ & 35 \\
NFGS            & $0.005\,\pm\,0.16$     &  $-0.214\,\pm\,0.57$ & 26 \\
BCD             & $-0.048\,\pm\,0.21$    &  $0.211\,\pm\,0.29$ & 17 \\
DGS             & $0.062\,\pm\,0.05$     &  $0.115\,\pm\,0.30$ & 13 \\
Virgo SFDs      & $0.103\,\pm\,0.09$     &  $0.078\,\pm\,0.62$ &  7 \\
Extra           & $0.069\,\pm\,0.05$     &  $0.107\,\pm\,0.12$ &  7 \\
\\
\hline 
\end{tabular}
}
\begin{flushleft}
{\footnotesize
$^{\mathrm a}$~The values given in Columns (2) and (3) are the means
and standard deviations of the logarithmic differences.
}
\end{flushleft}
\end{table}

\subsubsection{SFR}

Possibly the most difficult parameter is the SFR; the parent samples of MAGMA infer 
SFR originally using many different
methods, ranging from extinction-corrected \ha\ \citep[e.g., ALLSMOG, NFGS:][]{Cicone2017,Kannappan2013},
to hybrid FUV$+$IR or \ha$+$IR \citep[e.g., xGASS-CO, KINGFISH, BCDs, DGS, Virgo SFDs, ``Extra'':][]{Saintonge2017,Hunt2019,Hunt2016a,Grossi2015}
to the mean of different methods \citep[e.g., HRS:][]{Boselli2015}.
To calculate SFR for MAGMA, we have adopted the hybrid formulations of \citet{Leroy2019} based
on linear combinations of GALEX and WISE luminosities, estimated for their sample of 15\,750 galaxies within distances of $\sim$50\,Mpc.
Their expressions \citep[see Table 7 in][]{Leroy2019} are calibrated on the
GALEX-SDSS-WISE Legacy Catalogue (GSWLC) by \citet{Salim2016,Salim2018}, which were,
in turn, obtained by integrated population synthesis modeling relying on
CIGALE fits to $\sim$700\,000 low-redshift galaxies.
Thus, they are consistent with, and on the same scale, as our CIGALE-calibrated stellar masses.
Here we have converted their \citet{Kroupa2001} IMF to the \citet{Chabrier2003} one used here,
according to \citet{Speagle2014}.

\citet{Leroy2019} give several ``recipes'' for SFR in hybrid combinations:
we have preferentially used the expression with the smallest scatter,
namely luminosities of GALEX FUV combined with WISE W4.
The W4 luminosities are calculated in the same apertures as the W1 luminosities used for \mstar.
For GALEX, we adopted the magnitudes in the Revised catalog of GALEX UV sources by \citet{Bianchi2017}
that correspond to integrated values within elliptical apertures, 
and checked to make sure that the aperture size was commensurate with the WISE apertures. 
As for the \mstar\ estimates, the GALEX and WISE luminosities are corrected for
Galactic extinction using the \citet{Schlafly2011} $A_B$ values and the reddening curve from \citet{Draine2003}.
According to \citet{Leroy2019}, the FUV$+$W4 formulation gives a mean scatter of $\sim$0.17\,dex in log(SFR) for 
the $\sim$16\,000 galaxies they analyzed. 
SFRs derived from FUV$+$W4 were available for 277 MAGMA galaxies ($\sim$71\% of the sample), but if not, 
we adopted NUV$+$W4 (available for 66 galaxies, $\sim$17\%), which gives a slightly higher scatter ($\sim$0.18\,dex).
Overall, these two formulations gave the lowest systematic uncertainties for SFR, and are available for $\sim$88\% of the MAGMA galaxies.
If GALEX was unavailable, we relied on W4 alone (for 32 galaxies, $\sim$9\% of the sample), or otherwise
on the hybrid recombination line (\ha) luminosity combined with 24\,\micron\ luminosities (for 10 galaxies) as prescribed by \citet{Calzetti2010}
or on the original SFR value (7 galaxies: 1 ALLSMOG, 1 NFGS, 1 Virgo SFD, Sextans\,A, DDO\,154, and regions of DDO\,53 and DDO\,73).
All SFRs were converted to the \citet{Chabrier2003} IMF. 

Overall, as shown in Fig. \ref{fig:compare} (right panel),
the original SFRs and the new values agree reasonably well, with small
mean differences, and always zero to within the scatter (see Table \ref{tab:compare}).
The agreement with the original SFRs from xGASS-CO and HRS is particularly good, with virtually zero offsets
and scatters of $\sim$0.2\,dex.
Both of these samples derived SFR using hybrid schemes, not unlike the ones reported by \citet{Leroy2019} used here.
NFGS and the Virgo SFDs show the largest scatters, and for NFGS we attribute this to their use of \ha\ luminosities only,
corrected for extinction \citep[see][]{Kewley2005}.
The original SFRs for the Virgo SFDs were derived following \citet{Wen2014}, but the scatter is dominated by the galaxies with the lowest SFRs (and \mstar);
this effect for low-mass, low-metallicity dwarfs was also noted by \citet{Wen2014}, so is not unexpected.

\subsubsection{Metallicity}

As mentioned above, all gas-phase metallicities in our combined sample are either direct \te\
methods or calibrated through the \nii-based PP04N2 \citet{Pettini2004} calibration. 
When the original O/H calibration is not PP04N2,  
we have converted it to PP04N2 according to \citet{Kewley2008}.
The PP04N2 calibration has been shown to be the most consistent with \te\ methods \citep[see also][]{Hunt2016a}.
Extinction corrections for this calibration are very small because
the lines are very close in wavelength: $\lambda$\nii(a)\,=\,6549.86\,\AA, 
$\lambda$\nii(b)\,=\,6585.27\,\AA, \ha\,=\,6564.614\,\AA, so the extinction correction is negligible, for a 
\citet{Cardelli1989} extinction curve. Even for visual extinction \av\,=\,5\,mag, the relative correction is of the order of 1\%. 
This is well within the signal-to-noise of the \ha\ flux itself.

Since metallicity is found to decline from galactic centers to the peripheries 
\citep[e.g.,][]{Kewley2005,Pilyugin2014a,Pilyugin2014b,DeVis2019},
such gradients represent a possible source of systematics; 
therefore it is worth examining their impact on our metallicity estimates.
Some of the O/H for our sample are integrated \citep[e.g., NFGS, HRS, xGASS-CO, ALLSMOG:][]{Kewley2005,Boselli2013,Saintonge2017}, 
some are ``representative'', evaluated at 0.4\,R$_{\rm opt}$ \citep[e.g., KINGFISH:][]{Moustakas2010}, and some are nuclear
(e.g., the dwarf samples: BCDs, DGS, Virgo SFDs, and ``extra" sources).
\citet{Kewley2005} have quantified the difference among the global metallicities and the ones measured in
the nuclear regions for a sample of 101 star-forming galaxies selected from the NFGS. 
Independently of the galaxy type, they find that such difference amounts to $\sim 0.1$\,dex, 
a value which is similar to the typical statistical uncertainty of metallicities of 0.1--0.2\,dex. 
Metallicity gradients in late-type dwarf irregulars or BCDs are generally negligible
\citep[e.g.,][]{Croxall2009} or at most comparable to those in more massive spirals
\citep[e.g.,][]{Pilyugin2015}.
Thus, we conclude that metallicity gradients should not markedly affect our conclusions.

\begin{figure*}[!ht]
	\begin{minipage}{0.45\textwidth}
	\includegraphics[width=\textwidth]{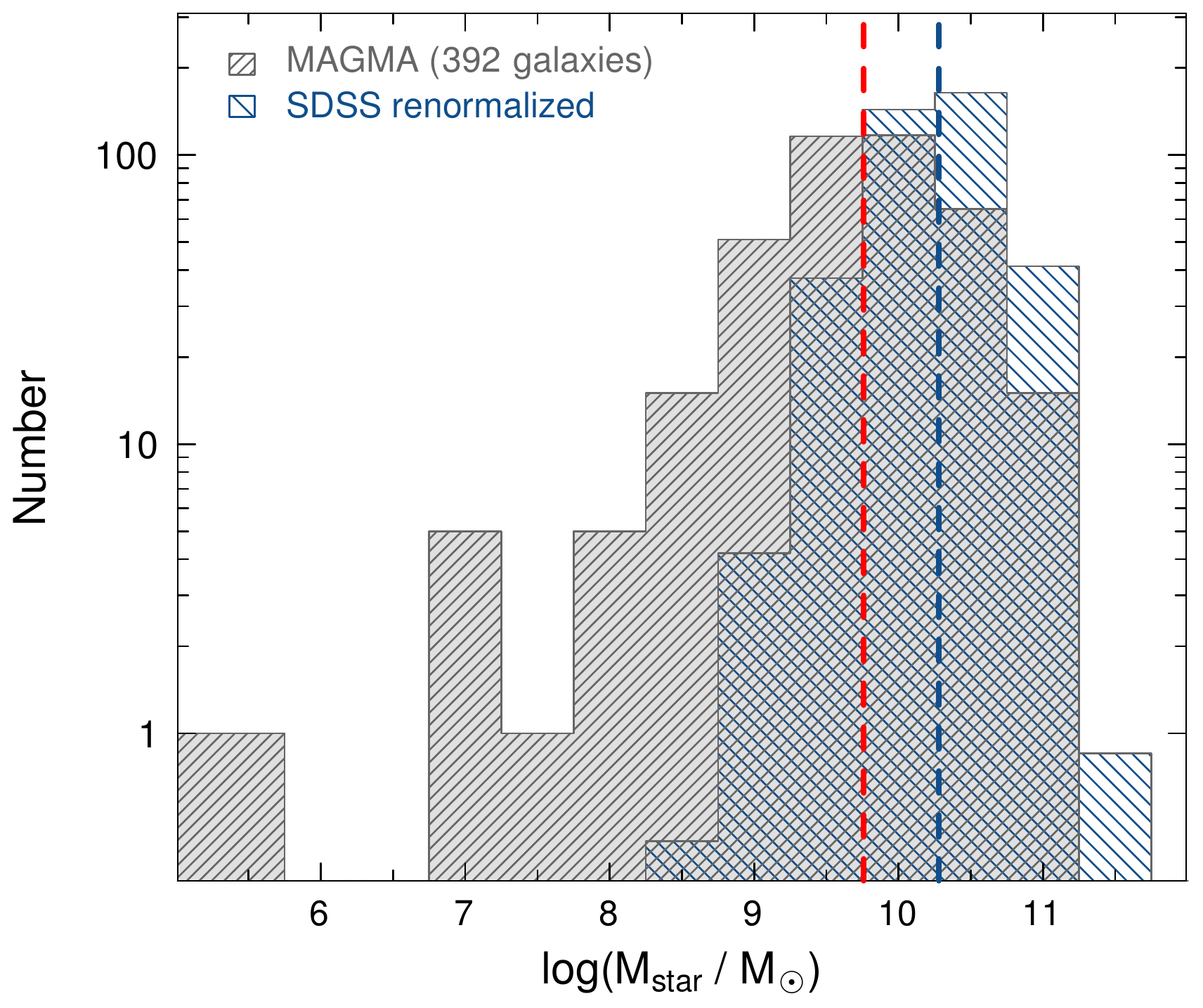}
	\end{minipage}
	\hspace{0.005\textwidth}
	\begin{minipage}{0.52\textwidth}
	\includegraphics[width=\textwidth]{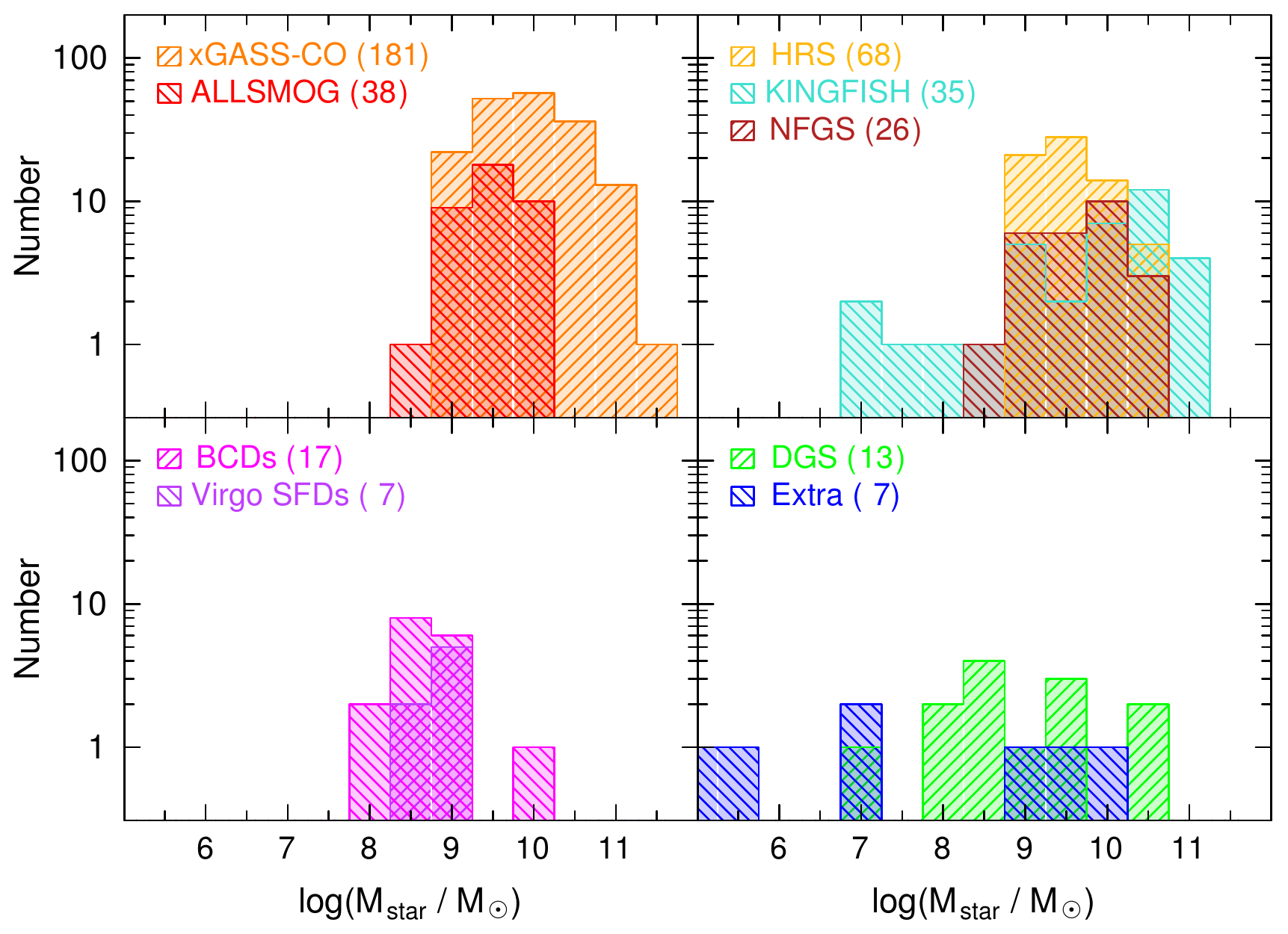}
	\end{minipage}
	\caption{Distributions of \mstar\ for the MAGMA sample; the right panels illustrate the subdivisions in \mstar\ for the parent surveys.
The red vertical dashed line in the left panel corresponds to the median \mstar\ for MAGMA, log(\mstar/\msun)\,=\,9.68.
Also shown in the left panel is the SDSS10 sample, taken from \citet{Mannucci2010}, consisting of 78579 galaxies. 
Here it has been renormalized to show the \mstar\ distribution it would have
if it contained the same number of galaxies (390) as MAGMA; the \mstar\ median of SDSS10 (shown as a blue vertical dashed line) is log(\mstar/\msun)\,=\,10.28. 
\label{fig:histogram_mstar}
}
\vspace{\baselineskip}
\setcounter{figure}{2}
	\begin{minipage}{\textwidth}
	\includegraphics[width=\textwidth]{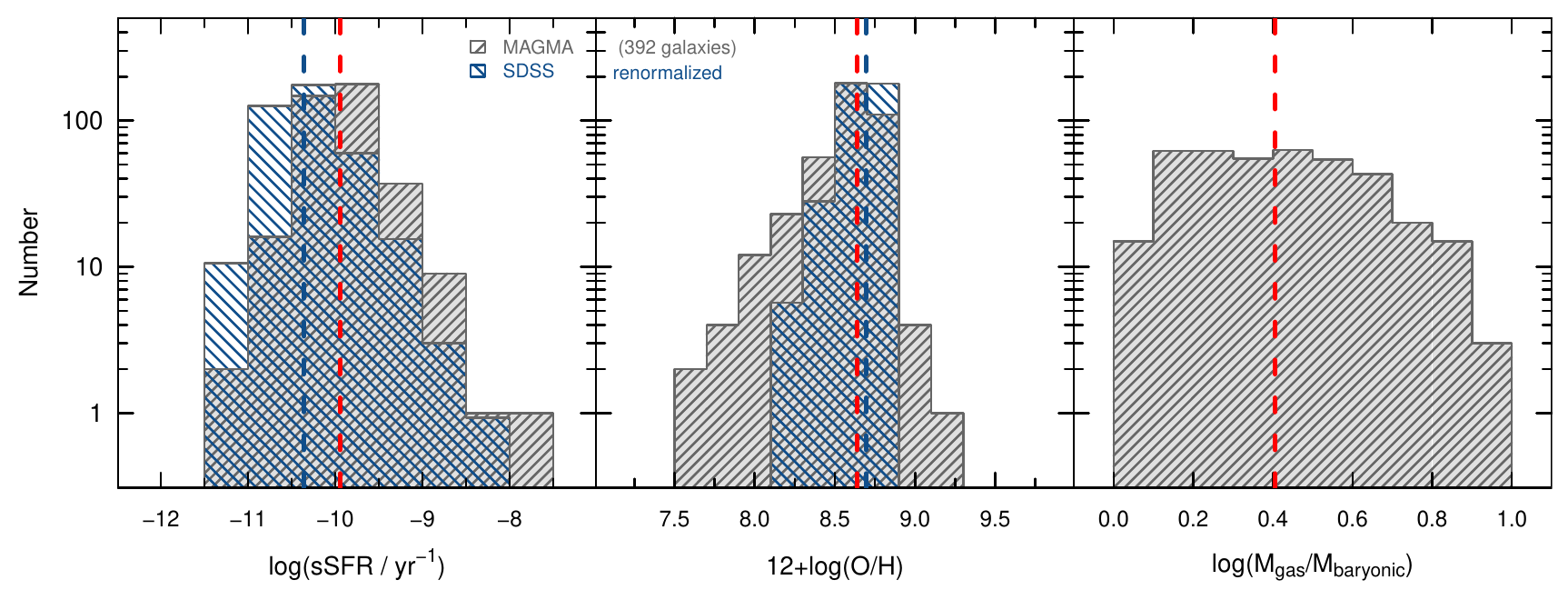}
	\caption{Distributions for combined sample of 392 galaxies for log\,(sSFR) (left panel), \logoh\ (middle), 
and log\,[\mgas/(\mstar\,$+$\,\mgas)\,$\equiv$\,\mgas/$M_{\rm baryonic}$] (right).  
The red vertical dashed lines in each panel correspond to the MAGMA medians: 
log(sSFR/yr$^{-1}$)\,=\,$-9.92$;
\logoh\,=\,$8.64$ \citep[\zsun\ would be \logoh\,=\,8.69, see][]{Asplund2009};
log(\mgas/M$_{\rm baryonic}$)\,=\,log(f$_{\rm baryonic}$)\,=\,$0.39$.
The left and middle panels also include the SDSS10 sample, as in Fig. \ref{fig:histogram_mstar}; 
the sSFR and O/H medians of SDSS10 are log(sSFR/yr$^{-1}$)\,=\,$-10.36$, and \logoh\,=\,8.69, respectively, shown by blue vertical dashed lines. 
}
	\label{fig:histogram_params}
	\end{minipage}
\end{figure*}

\subsubsection{Molecular gas mass}

Like metallicity, molecular gas mass is another delicate issue. 
Except for ALLSMOG and some KINGFISH galaxies, we use only CO(1--0) in order to avoid excitation issues; 
as mentioned before, to convert
to convert the CO(2--1) values to CO(1--0), a ratio of $R_{21}\,=\,0.8$ was assumed \citep[see also][]{Leroy2009}.
Here, the molecular gas masses have been calculated from \lco, using the conversion \mhtwo\,=\,\lco\,\aco\
(where \aco\ is the \htwo\ mass-to-CO light conversion factor), and adopting a metallicity-dependent calibration, following \cite{Hunt2015}. 
Specifically, for galaxies with $Z/Z_{\odot}<1$ \citep[i.e., \logoh$<$8.69, see][]{Asplund2009}, 
we applied $\alpha_{\rm CO} = \alpha_{\rm CO_\odot}  (Z/Z_{\odot})^{-1.96}$; 
for metallicities  $Z/Z_{\odot}\geq1$ we used a constant solar value of 
$\alpha_{\rm CO} = \alpha_{\rm CO_\odot}\,=\,3.2\,$\msun\,\kkmspc.

As mentioned previously, for MAGMA O/H, we have adopted either \te\ or the PP04N2 metallicity calibrations.
However, for the calculation of \aco, we have also investigated the effect of adopting an alternative strong-line calibration, 
namely the formulation by \citet[][KD02]{Kewley2002}.
To emulate \citet{Bothwell2016a,Bothwell2016b}, we
have also explored the \aco\ formulation from \citet{Wolfire2010} using KD02 metallicities.
This assumes that the \aco\ varies exponentially with visual extinction, \av, with a weak metallicity
dependence for \av\ \citep[see][for more details]{Bolatto2013}.
These results will be discussed in Sect. \ref{sec:bothwellcalibration}.

\begin{figure*}[!ht!]
	\centering
	\includegraphics[width=\textwidth]{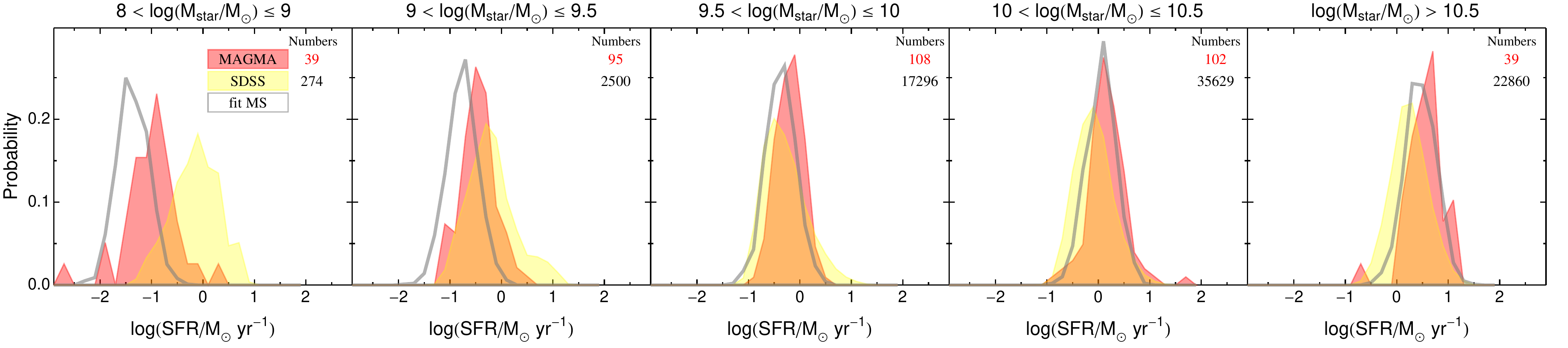}
	\caption{Distributions of SFR for the MAGMA sample shown as red-shaded regions in bins of \mstar;
also shown are the SDSS10 sample in yellow-shaded regions, and the 
fit to the SFMS (shown as a solid grey curve) reported by \citet{Hunt2019}.
The close agreement of MAGMA with SDSS and the SFMS, except possibly for the lowest \mstar\ bin, suggests that starbursts are not
dominating the galaxy population represented.
}
	\label{fig:sfr_distributions}
\end{figure*}

Aperture corrections for CO single-dish measurements are also a source of uncertainty,
because of the beam size compared to the dimensions of the galaxy.
Because of the relatively large distances of the xGASS-CO galaxies (0.025 $< z <$ 0.05 for \mstar$\geq 10^{10}$\,\msun), 
the median correction to total CO flux applied by \citet{Saintonge2017} is fairly small, a factor of 1.17.
Aperture corrections for the ALLSMOG galaxies \citep{Cicone2017} are even smaller, corresponding to a median
covering fraction of 0.98.

We have compared the CO luminosities \lco\ of the six galaxies common to HRS and KINGFISH.
This is an interesting comparison because the KINGFISH galaxies were mapped in CO(2--1) with the HERA CO Line Extragalactic Survey 
\citep[HERACLES,][]{Leroy2009}, and the HRS measurements were mostly single-dish CO(1--0) observations with few maps
\citep[][]{Boselli2014a}.
The mean difference between the two datasets is 0.04\,dex (larger luminosities for the HRS sample), 
with a standard deviation of 0.24\,dex.
Thus there is no systematic difference in \lco\ between these two samples, and, moreover, the spread
is similar to the typical aperture corrections of $\sim$0.2\,dex or less for xGASS-CO and ALLSMOG.

\subsubsection{Overall properties}

Finally, we compare median differences of each parent sample within MAGMA, relative to
the sample as a whole.
This is done in some detail in Appendix \ref{app:comparison} where we show this comparison graphically.
Our analysis shows that ultimately there are no significant systematic differences among the individual parent
samples, despite the different criteria for their original selection.
We therefore expect that MAGMA, as a whole, is representative of field galaxies in the Local Universe,
and can be used to assess the gas scaling relations driving baryonic cycling.  

The \mstar\ distributions of our combined sample are shown in Fig. \ref{fig:histogram_mstar}, 
while sSFR, $Z$ [measured in units of \logoh] 
gas fractions \mgas/(\mgas\ + \mstar) distributions are shown in Fig. \ref{fig:histogram_params}. 
The combined MAGMA sample  
covers the following unprecedented ranges in parameter space,
spanning more than 5 orders of magnitude in \mstar, SFR, and \mgas, 
and almost 2 orders of magnitude
in metallicity\footnote{Here and elsewhere throughout this paper,
``log'' means decimal logarithm unless otherwise noted.}: 
\begin{equation}
\begin{gathered}
5.2 \lesssim \log({\rm M}_{\rm star}/M_{\odot}) \lesssim 11.2 \\ \nonumber
	5.4 < \log({\rm M}_{\rm gas}/M_{\odot}) < 10.8 \\ \nonumber
	-4.6 \lesssim \log({\rm SFR}/{\rm M}_{\odot}~{\rm yr^{-1}}) \lesssim 1.6 \\ \nonumber
	7.5 \lesssim 12 + \log{\rm (O/H)} \lesssim 9.3 
\end{gathered}
\end{equation}

To demonstrate the general applicability of results obtained for MAGMA to the general (field) galaxy population
in the Local Universe, we have included in Figs. \ref{fig:histogram_mstar} and \ref{fig:histogram_params}
the parameter distributions for the SDSS-DR7 catalogue consisting of 
$\sim$79000 galaxies from \citet{Mannucci2010};
hereafter we refer to this sample as SDSS10.
For a consistent comparison with MAGMA, like for the samples described above,
we have transformed the original O/H calibration from \citet{Maiolino2008} based on KD02 to PP04N2 according
to the formulation of \citet{Kewley2008} \citep[for more details, see also][]{Hunt2016a};
according to \citet{Kewley2008}, this transformation has an accuracy of $\sim$0.05\,dex.  
The distributions shown in Figs. \ref{fig:histogram_mstar} and \ref{fig:histogram_params} have been renormalized to
the number of galaxies in the MAGMA sample, to be able to compare the number distributions directly.
The SDSS10 has a relatively narrow spread in O/H; there are 5 MAGMA galaxies ($\sim 1$\%) 
beyond the highest metallicities in SDSS10, and all are from the xGASS-CO sample,
corresponding also to some of the most massive galaxies. 
The MAGMA \mstar\ median is 0.6\,dex lower than for the SDSS10, 
and the sSFR median (Fig. \ref{fig:histogram_params}) is $\sim$0.4\,dex higher,
illustrating that the MAGMA sample contains more low-mass galaxies than SDSS10.
Interestingly, there are only 25 SDSS10 galaxies (0.03\%) with log(\mstar/\msun)$>$11.5; thus because of
the normalization in Fig. \ref{fig:histogram_mstar} they do not appear. 

Both MAGMA and SDSS10 contain a large percentage of massive galaxies relative to local volume-limited samples
such as the Local Volume Legacy \citep[e.g., LVL,][]{Dale2009,Kennicutt2009}
or galaxy-stellar mass functions (GSMF).
For the $z \sim 0$ GSMF determined by \citet{Baldry2012} we would expect $\sim$0.1\% of the galaxies to be more
massive than the break mass, M$_*\,=\,5\times10^{10}$;
$\sim$14\% of the galaxies in SDSS10 and $\sim$5\% of those in MAGMA are more massive than this.
The preponderance of massive galaxies in these two samples, relative to a volume-limited one, is due
to flux limits, and the necessity of ensuring spectroscopic measurements (for SDSS10) and
CO measurements (for MAGMA). 
In any case, as shown in Figs. \ref{fig:histogram_mstar} and \ref{fig:histogram_params},
the parameter coverage of MAGMA does not deviate significantly from SDSS10 at high \mstar\ and O/H, 
but substantially extends the parameter space to lower \mstar\ and O/H values.

Given that we required that CO be {\it detected}, even at low metallicity, there could be a chance that the
MAGMA sample is dominated by starbursts, i.e., galaxies with SFRs significantly above the main sequence.
We examine this possibility in Fig. \ref{fig:sfr_distributions},
where we compare the distributions for MAGMA galaxies of SFR in different bins of \mstar\ with
the SDSS10 data from \citet{Mannucci2010} as above. 
Also shown is the main sequence of star formation given by fitting LVL and KINGFISH by \citet{Hunt2019},
here approximated by a Gaussian distribution with a width of 0.3\,dex
\citep[see also][]{Renzini2015}.
Except for possibly the lowest-mass bins, log(\mstar/\msun)\,$\leq$\,9, Fig. \ref{fig:sfr_distributions} demonstrates
that the MAGMA sample is well approximated by main-sequence SFR distributions.
Thus, it is not dominated by starbursts, and can be considered a reliable diagnostic for gas processes
in the Local Universe.

\begin{figure}[!t]
	\centering
	\includegraphics[width=0.49\textwidth]{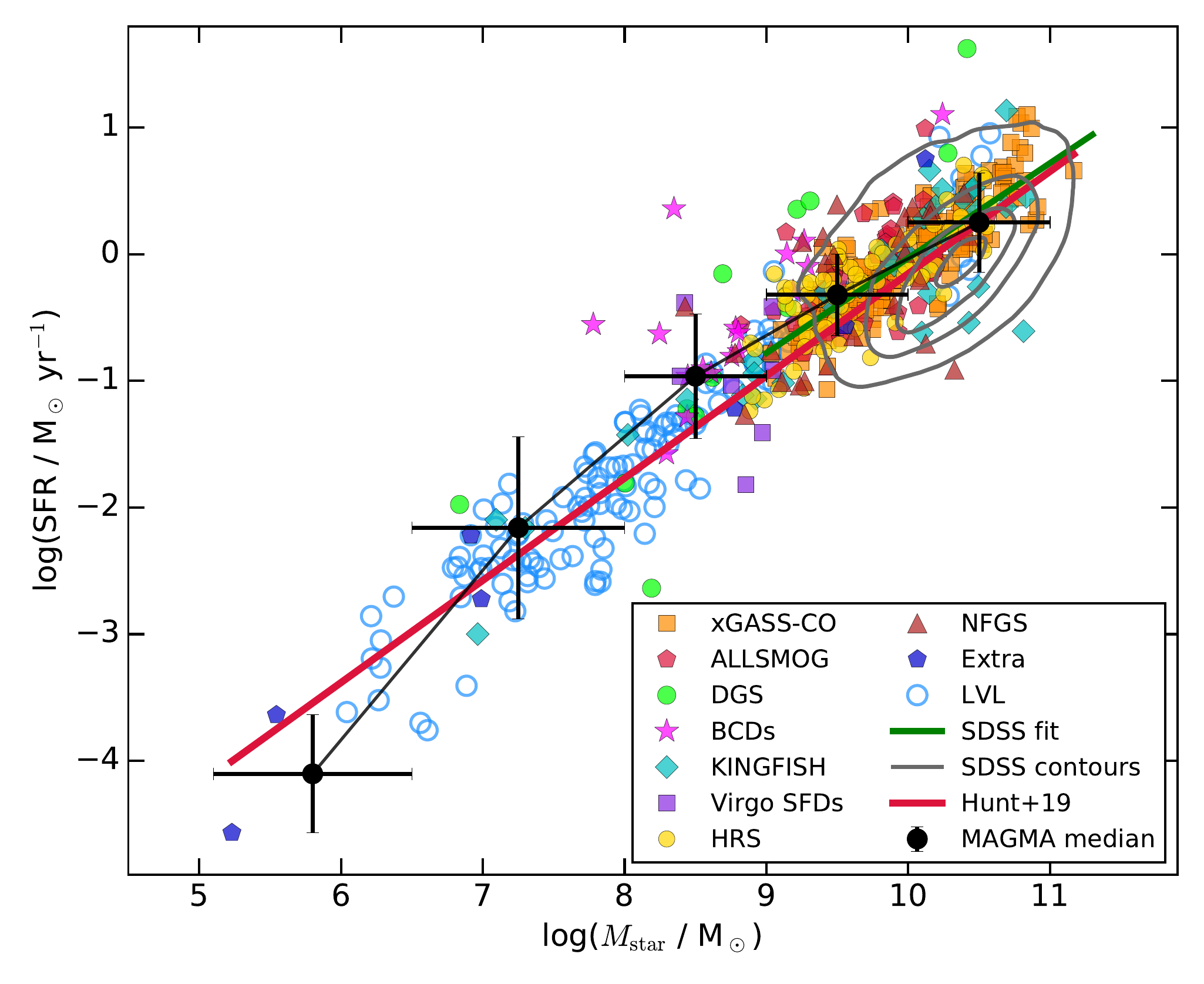} \\
\hspace{0.02\textwidth}
	\includegraphics[width=0.49\textwidth]{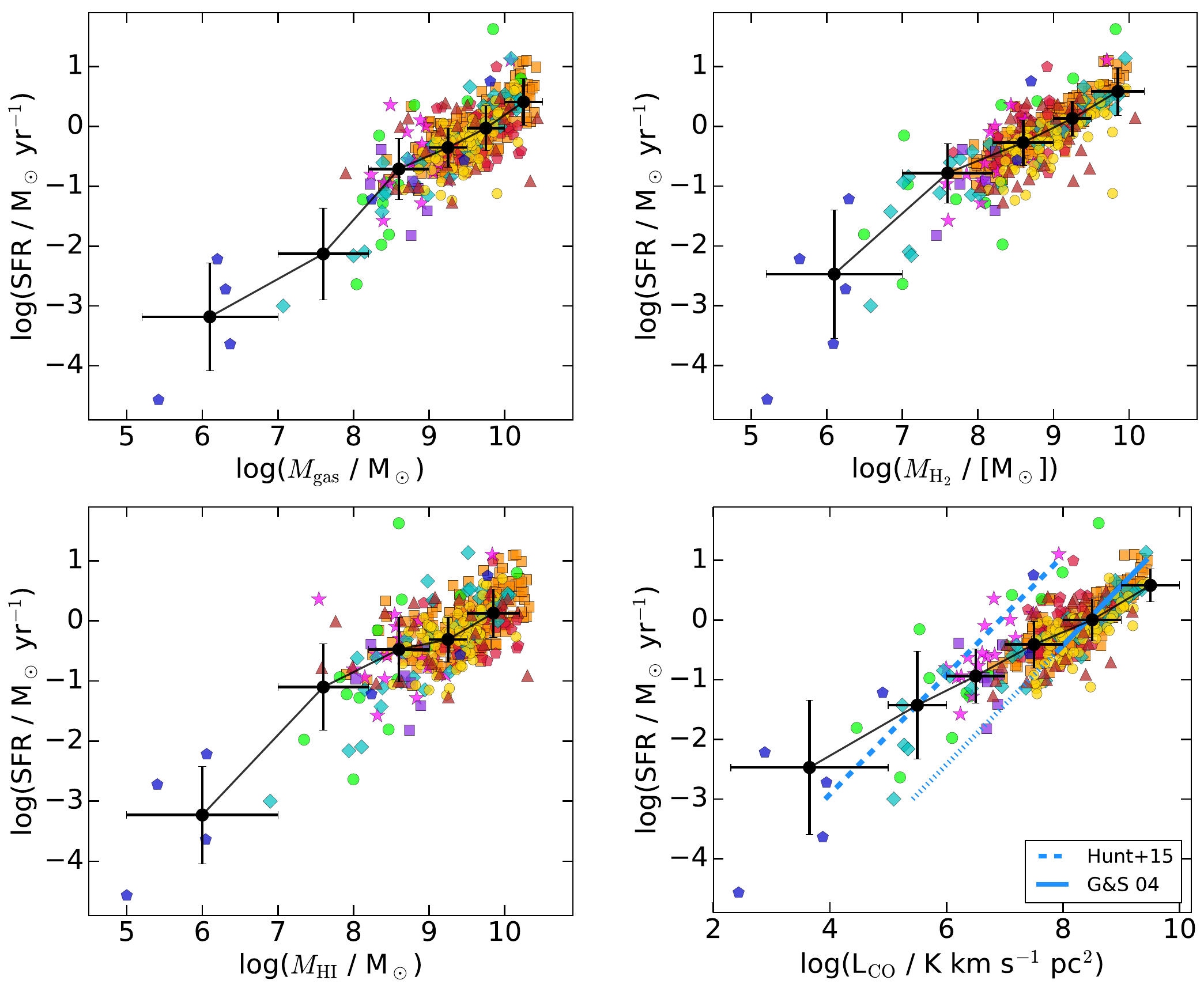}
	\caption{Upper panel: Galaxies from the MAGMA sample plotted in the SFR--\mstar\ plane. 
		The 
green line is the MS of star-forming galaxies derived from the SDSS \citep{Brinchmann2004}.
Contours in this plane for the SDSS10 sample are shown as grey closed curves;
the lowest-level contour encloses 90\% of the sample.
Also shown in the upper panels is the LVL sample \citep{Dale2009,Kennicutt2009} as described in the text.
		Lower panels: MAGMA galaxies in the SFR--\mgas\ (upper-left), 
\mhtwo--SFR (upper-right), \mhi--SFR (lower-left), and \lco--SFR (lower-right) planes.
		Symbols and colours refer to different parent surveys, as
indicated in the legend. 
In all panels, the grey lines represent the median trends of
the MAGMA distributions, calculated at different bins (see black dots; horizontal
bars indicate the widths of the bins, while vertical bars indicate the standard
deviation around median values of galaxies in the bins). 
In the lower right panel, the blue solid line is the fit relating \lco\
and SFR from \citet{Gao2004},
and the dotted line roughly parallel to it but offset by roughly a factor of 30 is
the analogous fit for low-metallicity dwarf galaxies by \citet{Hunt2015}.
The dotted extension of the regression found by \citet{Gao2004}
reflects the range of parameters for which they calibrated the relation.
	\label{fig:scaling1}
}
\end{figure}

\begin{figure}[!h]
	\centering
	\includegraphics[width=0.9\columnwidth]{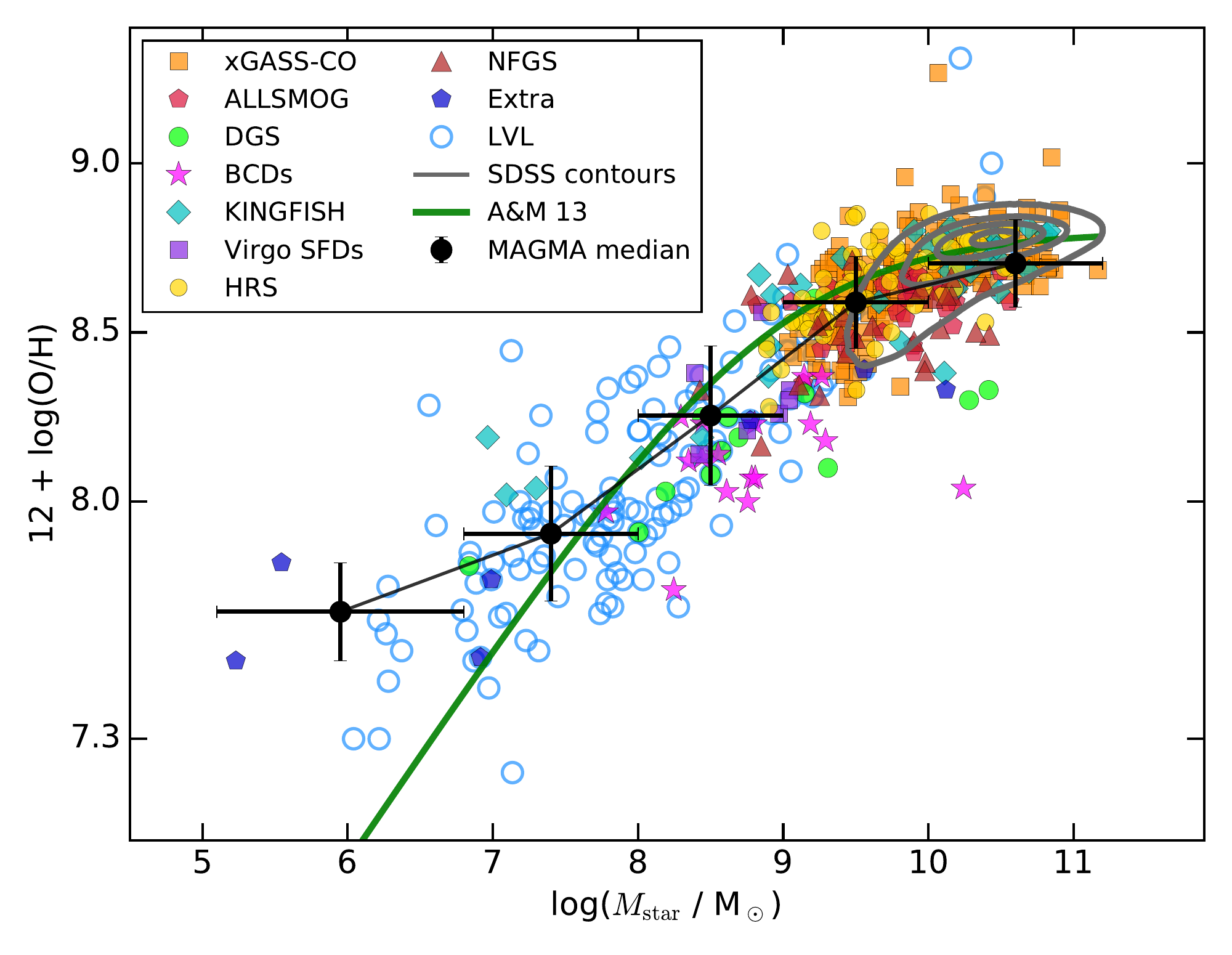} \\
	\includegraphics[width=0.95\columnwidth]{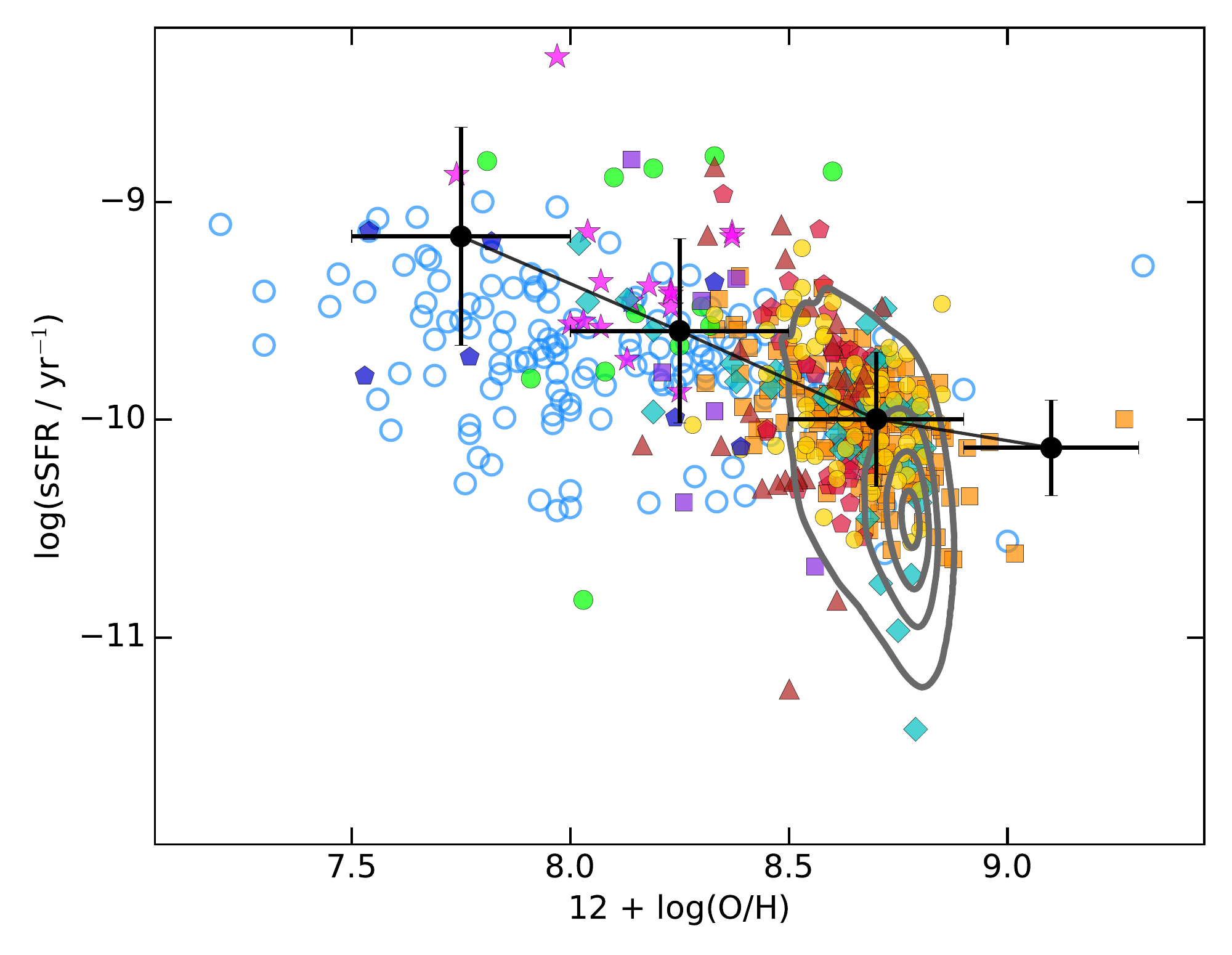} 
	\caption{Upper panel: the MAGMA sample in the \mstar--$Z$ plane;
lower panel: MAGMA sample in the sSFR--Z plane. Metallicities are reported in PP04N2 (or \te) units of \logoh.
In both panels, as in previous figures, the SDSS10 sample is also plotted,
with 90\% of the galaxies enclosed within the lowest contours;
both panels show also the LVL sample \citep{Dale2009,Kennicutt2009} as described in the text.
In the upper panel, the MZR from the direct-\te\ determination by \citet{Andrews2013} is
also given, and its good agreement with the MAGMA PP04N2 ($+$\te) metallicities suggests that
the calibrations are consistent.
Symbols and colours refer to different parent surveys, as
indicated in the legend. The black dashed line represents the median trend of
the distribution, calculated at different bins (see black dots; horizontal bars
indicate the widths of the bins, while vertical bars indicate the standard
deviation around median values of galaxies in the bins).
\label{fig:mzr}
}
\end{figure}

\subsection{Fundamental scaling relations}

In Fig. \ref{fig:scaling1}, MAGMA galaxies are plotted in the \mstar--SFR plane, forming the 
SFMS;
the lower panels of Fig. \ref{fig:scaling1} show various forms of the correlation between SFR and \mgas, 
a global SK law, exploring different gas phases 
(atomic, molecular and total, i.e., \mhi$+$\mhtwo) and CO luminosity, \lco.
Also shown in Fig. \ref{fig:scaling1} are the loci of the SDSS10 galaxies reported in previous figures.
The lowest-level contour encloses 90\% of the sample, illustrating the extension by MAGMA to lower \mstar\ and SFR.
We have included in Fig. \ref{fig:scaling1} also the parameters from the LVL sample as measured by \citet{Hunt2016a};
most metallicities are from the direct \te\ method \citep{Marble2010,Berg2012}.
This sample is the best approximation available for the number and types of galaxies present in the nearby Universe.

Figure \ref{fig:mzr} shows the mass-metallicity (\mstar--$Z$) relation of the MAGMA sample; 
although with some scatter, galaxies lie along the MZR over $>4$ dex in \mstar\ and 
a factor of $\sim 50$ in $Z$ (1.7\,dex in \logoh).
As in Fig. \ref{fig:scaling1}, we have also included SDSS10 and LVL;
the MAGMA sample is consistent with both, implying that there are no significant selection effects
from our gas detection requirements.
Also shown as a blue curve in Fig. \ref{fig:mzr} is the direct-\te\ calibration for the SDSS
obtained by \citet{Andrews2013}; the MAGMA PP04N2$+$direct \te\ metallicities are in good agreement
with this calibration illustrating that PP04N2 is a good approximation to \te\ methods \citep[see also][]{Hunt2016a}.

Fig. \ref{fig:mzr} illustrates that the flattening of the MZR that frequently emerges at high \mstar\
is present in the MAGMA and SDSS10 samples.
This curvature is clearly seen in the contours of SDSS10 where 90\% of the SDSS10 galaxies
are enclosed in the lowest contour.
Again, the overall extension of MAGMA to lower \mstar\ and O/H is evident.
In what follows, we focus on applying linear scaling relations to this curved MZR.
A single linear relation is not altogether appropriate, given the flattening of the MZR at high \mstar.
Indeed, as shown by the linear trend in the upper panel of Fig. \ref{fig:mzr}, it can only roughly approximate the overall MAGMA distribution.
In what follows, we investigate the best approach to approximate non-linear trends in the data.

A combination of the scaling relations described above produce the result reported in 
the lower panel of Fig. \ref{fig:mzr}, namely a tight correlation between the specific SFR (sSFR = SFR/\mstar) and 
metallicity over $\sim 2$ dex in sSFR and $\ga 1.5$ dex in \logoh. 
Fig. \ref{fig:mzr} demonstrates that galaxies more actively forming stars (i.e., with a high sSFR) are less enriched
\citep[and also more gas-rich; see][for a discussion]{Mannucci2010,Cresci2012,Hunt2016b,Cresci2018}.

In the next section we focus on the MZR (Fig. \ref{fig:mzr}), exploring its secondary and tertiary dependencies on SFR and \mhtwo, \mhi, and \mgas.
Since $Z$ is the only {\it intensive} 
\footnote{Intensive properties are physical properties of a system that do not depend on the system size or the amount of material in the system 
(e.g., the metallicity of a galaxy does not depend on its size). They differ from 
{\it extensive} properties, that are additive for subsystems. For instance, the total $M_{\rm star}$, $M_{\rm gas}$ and SFR 
in a galaxy are the sums of the parts composing the galaxy; in other words these quantities depend on the size of a galaxy.} %
quantity among the available integrated physical properties of our galaxies, the MZR is, among the others described above, 
the most sensitive to  internal physical mechanisms.

\section{Mutual correlations: a PCA analysis \label{sec:pca}}

Principal Component Analysis (PCA) is a parameter transformation technique that diagonalizes the covariance matrix of a set of variables. 
Consequently, a PCA gives the linear combinations of observables, the eigenvectors, that 
define the orientations whose projections constitute a hyper-plane; 
these eigenvectors minimize the covariance and are, by definition, mutually orthogonal.
In other words, a PCA performs a coordinate transformation that identifies the optimum 
projection of a dataset and the parameters that are most responsible for the variance in the sample.
The most common use of PCA is to search for possible 
{\it dimensionality reduction} of the parameter space needed to describe a dataset. 
For instance, a PCA approach has shown that galaxies lie roughly on a 2D surface in the 3D parameter space defined by \mstar, SFR and $Z$ 
\citep[e.g.,][]{Hunt2012,Hunt2016a} or \mstar, \mhtwo\ and $Z$ \citep[e.g.,][]{Bothwell2016a}.

We use the MAGMA sample to expand upon and re-examine previous trends found with 
PCAs of \mstar, SFR, metallicity and gas.
In addition to the ``classical'' algorithm for PCA (an unweighted matrix diagonalization),
we also apply two additional PCA methods which give the uncertainties associated with the best-fit parameters:
the ``bootstrap PCA'' (BSPCA) first introduced by \citet{Efron1979,Efron1982} 
and the ``probablistic PCA'' (PPCA) described by \citet{Roweis1998}. 
BSPCA is a specific example of more generic techniques that resample the original data set with replacement, 
to construct ``independent and identically distributed'' observations. 
PPCA is an expectation-maximization (EM) algorithm which also accommodates missing information.
For the PPCA, we randomly remove 5\% of the individual entries for each galaxy; in practice, this
means that we omit the SFR for 5\% of the galaxies, \mstar\ for a different 5\%, and so on.
For both methods, we generate several realizations of 100--1000 repetitions, and calculate the means and standard deviations 
of the resulting PC coefficients.
For all statistical analysis, we rely on the
{\it R} statistical package\footnote{{\it R} is a free software environment for statistical 
computing and graphics (\url{https://www.r-project.org/}).}.

To estimate uncertainties,
other groups have used Monte Carlo methods with resampling, injecting Gaussian noise into the nominal measurement values
\citep[e.g.,][]{Bothwell2016a,Bothwell2016b}.
We performed several detailed tests using this technique and found that it introduces systematics in the
results, depending on the amplitude of the noise and the \mstar\ and SFR distributions of the samples;
hence we prefer resampling techniques in order to avoid potential unreliability of the results.
This point will be discussed further in Sect. \ref{sec:methods} and Appendix \ref{app:injection}.

\begin{table*}[!th!] 
\caption{4D PCA results for MAGMA$^{\mathrm a}$}
\resizebox{\linewidth}{!}{
\label{tab:4dpca}
\begin{tabular}{lcccccccc}
\hline 
\\ 
\multicolumn{1}{c}{Method} &
\multicolumn{1}{c}{PC4(1)} &
\multicolumn{1}{c}{PC4(2)} &
\multicolumn{1}{c}{PC4(3)} &
\multicolumn{1}{c}{PC4(4)} &
\multicolumn{1}{c}{PC4} &
\multicolumn{1}{c}{PC4} &
\multicolumn{1}{c}{PC3} &
\multicolumn{1}{c}{PC1$+$PC2} \\
& \multicolumn{1}{c}{\logoh} &
\multicolumn{1}{c}{log} &
\multicolumn{1}{c}{log} &
\multicolumn{1}{c}{log($x$)} &
\multicolumn{1}{c}{std. dev.} &
\multicolumn{3}{c}{proportion of variance} \\
& \multicolumn{1}{c}{(PP04N2)} &
\multicolumn{1}{c}{(\mstar/\msun)} &
\multicolumn{1}{c}{(SFR/\msunyr)} \\ 
\multicolumn{1}{c}{(1)} &
\multicolumn{1}{c}{(2)} &
\multicolumn{1}{c}{(3)} &
\multicolumn{1}{c}{(4)} &
\multicolumn{1}{c}{(5)} &
\multicolumn{1}{c}{(6)} &
\multicolumn{1}{c}{(7)} &
\multicolumn{1}{c}{(8)} &
\multicolumn{1}{c}{(9)} \\
\\ 
\hline 
\\
&&&&\multicolumn{1}{c}{$x\,=\,$\mhi/\msun} \\
PCA  &   0.920 &   $-0.355$ &  0.164 & $0.019$ &  0.127 &  0.010 & 0.051 & 0.94\\
PPCA &   $0.95 \pm 0.01$ & $-0.29 \pm 0.03$  &  $0.11 \pm 0.03$  & $-0.01 \pm 0.01$  &  0.14  & 0.01  \\ 
BSPCA &  $0.92 \pm 0.01$ & $-0.36 \pm 0.03$  &  $0.16 \pm 0.03$  & $0.02 \pm 0.02$  &  0.13  & 0.01  \\ 
\\
&&&&\multicolumn{1}{c}{$x\,=\,$\mhtwo/\msun} \\
PCA   &   0.913            &   $-0.378$        &  0.133            & $0.079$ & 0.124 &  0.010 & 0.051 & 0.94\\
PPCA  &  $0.94 \pm 0.01$   & $-0.31 \pm 0.03$  &  $0.09 \pm 0.03$  & $0.05 \pm 0.02$ &  0.14   & 0.01  \\ 
BSPCA &  $0.91 \pm 0.02$   & $-0.38 \pm 0.03$  &  $0.13 \pm 0.03$  & $0.08 \pm 0.04$ &  0.12   & 0.01  \\ 
\\
&&&& \multicolumn{1}{c}{$x\,=\,$\mgas/\msun} \\
PCA   &  0.917           &  $-0.366$         &   0.153           & $0.048$         &  0.126 &  0.011 & 0.054 & 0.94\\
PPCA  &  $0.95 \pm 0.01$ & $-0.24 \pm 0.02$  &  $0.13 \pm 0.02$  & $-0.09 \pm 0.03$  &  0.14  & 0.02  \\ 
BSPCA &  $0.91 \pm 0.02$ & $-0.37 \pm 0.03$  &  $0.16 \pm 0.03$  & $0.05 \pm 0.03$  &  0.13  & 0.01  \\ 
\\
&&&& \multicolumn{1}{c}{$x\,=\,$\lco/\lcounits} \\
PCA   &  0.955            & $-0.185$         &   0.181           & $-0.147$          &  0.117 & 0.007 & 0.027 & 0.97\\
PPCA  &  $0.97 \pm 0.01$  & $-0.16 \pm 0.04$ &  $0.13 \pm 0.02$  & $-0.13 \pm 0.03$  &  0.13  & 0.01  \\ 
BSPCA &  $0.95 \pm 0.01$  & $-0.18 \pm 0.04$ &  $0.18 \pm 0.03$  & $-0.15 \pm 0.02$  &  0.12  & 0.01  \\ 
\\
\hline 
\\
\multicolumn{1}{c}{Method} &
\multicolumn{1}{c}{PC4(1)} &
\multicolumn{1}{c}{PC4(2)} &
\multicolumn{1}{c}{PC4(3)} &
\multicolumn{1}{c}{PC4(4)} &
\multicolumn{1}{c}{PC4} &
\multicolumn{1}{c}{PC4} &
\multicolumn{1}{c}{PC3} &
\multicolumn{1}{c}{PC1$+$PC2} \\
& \multicolumn{1}{c}{\logoh} &
\multicolumn{1}{c}{log} &
\multicolumn{1}{c}{log} &
\multicolumn{1}{c}{log} &
\multicolumn{1}{c}{std. dev.} &
\multicolumn{1}{c}{variance} &
\multicolumn{1}{c}{variance} &
\multicolumn{1}{c}{variance} \\
& \multicolumn{1}{c}{(KD02)} &
\multicolumn{1}{c}{(\mstar/\msun)} &
\multicolumn{1}{c}{(SFR/\msunyr)} & 
\multicolumn{1}{c}{(\mhtwo/\msun)$^{\mathrm b}$} \\
\\
\hline 
\\ 
PCA   &  $0.86$          & $-0.45$           &  $0.22$           & $0.03$           &  0.153 &  0.014 & 0.048 & 0.95\\
PPCA  &  $0.92 \pm 0.02$ & $-0.37 \pm 0.04$  &  $0.15 \pm 0.04$  & $0.02 \pm 0.02$  &  0.17  & 0.02  \\ 
BSPCA &  $0.86 \pm 0.03$ & $-0.45 \pm 0.04$  &  $0.22 \pm 0.04$  & $0.03 \pm 0.03$  &  0.15  & 0.01  \\ 
\\
\hline 
\end{tabular}
}
\begin{flushleft}
{\footnotesize
$^{\mathrm a}$~In PCA, the relative signs of the PCs are arbitrary, so that
we have used the same conventions for all; 
this has no bearing on the inversion of the equation of the PC with the least variance. 
Column (6) reports the standard deviation of PC4 around the hyperplane, and Cols. (7--9) give the proportion of 
sample variance contained in PC4, PC3, and the sum of PC1$+$PC2, respectively.
\\
$^{\mathrm b}$~Here \mhtwo\ is calculated with \aco\ according to the exponential formulation of \citet[][]{Wolfire2010,Bolatto2013},
using the KD02 metallicities.
}
\\
\end{flushleft}
\end{table*}


Thus, we perform three kinds of PCAs on: 
(1) a 4D parameter space defined by \mstar, SFR, $Z$, and a gas quantity (either \mgas, \mhtwo, \mhi, or \lco);
and (2) a 3D space defined by \mstar, SFR, and either metallicity $Z$ or a gas quantity (as for 4D).
We then assess whether two, three, or four parameters 
are statistically necessary to describe the variance of these quantities
in the MAGMA sample; this is decided by comparing the change in scatter produced by adding an additional variable.
\lco\ has been included as a separate gas quantity in order to separate the effects of true CO luminosity from the effects
of a metallicity-dependent \aco; this point will be discussed further below. 
We have also performed a five-dimensional PCA on \mstar, SFR, $Z$, \mhi, and \lco\ (or \mhtwo),
but results do not differ significantly from the 4D case, so we do not discuss it here. 

\bigskip
\subsection{4D PCA}
\label{sec:4dpca}


Results for the 4D PCA are given in Table \ref{tab:4dpca}; the coefficients of the PC with the least variance (by definition PC4) are reported,
together with the fraction of variance contained in PC4, PC3, and the sum of PC1$+$PC2. 
There are 
two separate rows for the PPCA and the BSPCA; these are the different methods
used to infer uncertainties, and demonstrate that the coefficients of all the methods agree to within the uncertainties. 
We find that the proportion of variance contained in only the first two eigenvectors, PC1$+$PC2, is 
generally large, $\ga$\,94\%.
Because most of the variance of the sample is contained within the first two eigenvectors, the dimensionality of the parameter space
of the MAGMA galaxies is two-fold.
They are distributed on a 2D plane in the 4D space; 
the remaining $\la$\,6\% of the variance (shared between PC3 and PC4) produces a scatter perpendicular to this plane.

PC4, the eigenvector with the least variance ($\sim$1\%), is always dominated by metallicity, $Z$
(see Table \ref{tab:4dpca}).
Because very little of the sample variance is contained in PC4,
it can be set to zero and inverted to give a useful prediction for the dominant term, namely metallicity $Z$ \citep[see][for a discussion]{Hunt2012,Hunt2016a}.
The estimate for the metallicity obtained by setting PC4 equal to zero is formally accurate to the 1--2\% level,
i.e., the variance associated with PC4; however, a more robust assessment of the accuracy is obtained by fitting the
residuals to a Gaussian as described below.

Interestingly, the contribution of \mhi\ to PC4 is consistent with zero within the uncertainties,
and the coefficients for \mhtwo\ and \mgas\ are small, determined to be non-zero only at a 2$\sigma$ level or less.
The PC4 coefficient for all gas components is smaller than that for SFR.
The only gas PC4 coefficient strongly different from 0 is \lco, determined at $\sim 5-7\sigma$,
and comparable in magnitude to the SFR coefficient.
This result implies that the 2D plane does not depend significantly on gas properties (except for possibly CO luminosity \lco).
The expression for \logoh\ obtained by inverting the expression based on \mhtwo\ is:
\begin{eqnarray}
o & = & (0.42\,\pm\,0.03)\,m - (0.15\,\pm\,0.04)\,s - (0.09\,\pm\,0.04)\,h_2 \nonumber \\
\label{eqn:pca4dh2}
\end{eqnarray}
\noindent
where $h_2$, $m$, $o$, and $s$ are defined as the centered variables, i.e., log(\mhtwo), log(\mstar), \logoh, and log(SFR) 
minus their respective means in the MAGMA sample as given in Table \ref{tab:means}.
The accuracy of this expression is $\sim$0.12\,dex, 
assessed by fitting a Gaussian to the residuals of this fit compared to the observations of \logoh. 
Eqn. (\ref{eqn:pca4dh2}), in which the uncertainties from Table \ref{tab:4dpca} have been propagated, tells us that:
\begin{itemize}
\renewcommand\labelitemi{--}
\item
the gas-phase metallicity of galaxies in the MAGMA sample is primarily dependent on \mstar\ (a confirmation of the existence of the well-known MZR);
\item
there is a strong secondary dependence on the SFR, whose contribution in determining $Z$ is $\sim$40\% as strong as the dependence on  \mstar;
\item
the tertiary dependence on \mhtwo\ is negligible, virtually zero, given that the accuracy with which the coefficient is determined is $\la 2\sigma$. 
\end{itemize}

\begin{figure*}[!ht]
\includegraphics[width=0.95\textwidth]{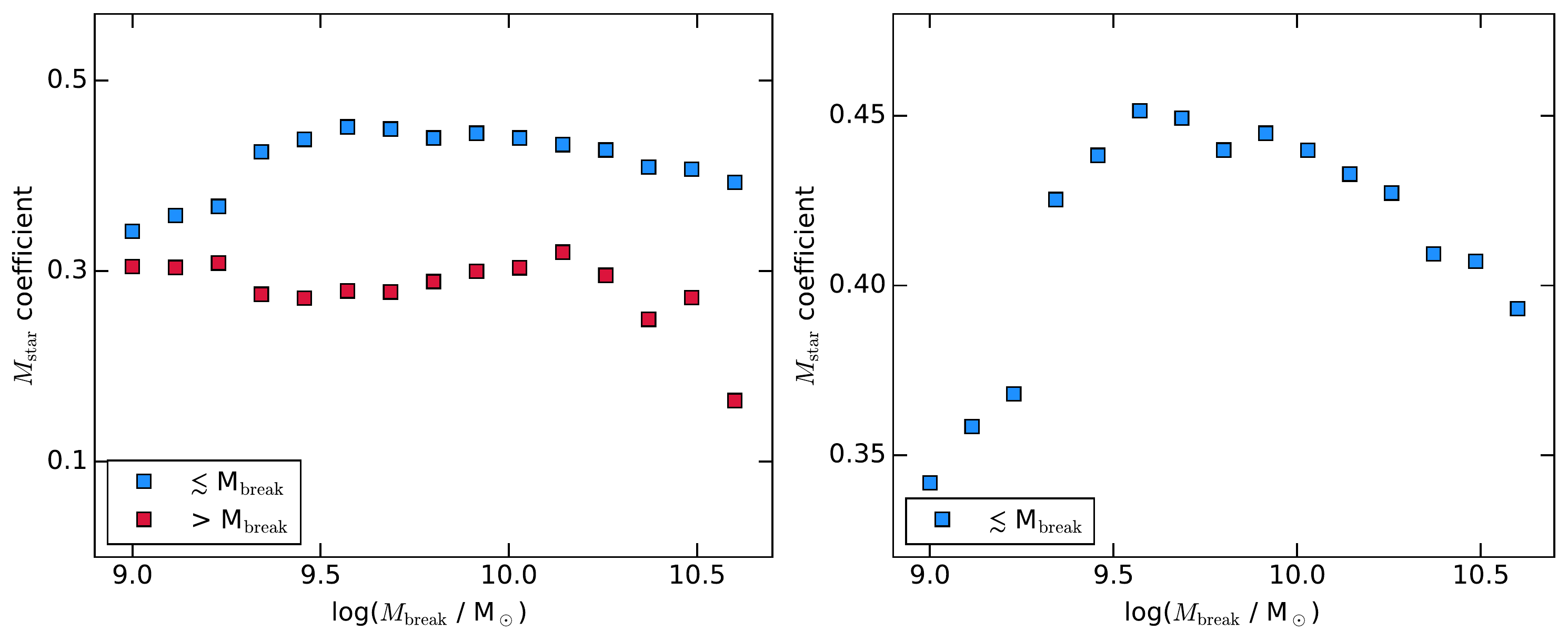}
\caption{PCA coefficients for MAGMA log(\mstar) plotted against
\mbreak\ for the division into the two PCAs.
The right panel is a magnification of the left panel, showing more effectively the range of \mstar\
coefficients as the \mbreak\ mass is varied.
The lower values of the \mstar\ coefficient toward higher \mbreak\ results from the flattening
of the MZR at high \mstar.
}
\label{fig:mstar_coeffs}
\end{figure*}

We have explored the behavior of the other gas quantities, and as suggested by
Table \ref{tab:4dpca}, it is similar to the behavior of \htwo; with the possible exception of \lco,
the gas content is not influencing metallicity $Z$.
However, in the case of \lco, the \mstar\ coefficient is significantly smaller than that 
for the other gas quantities and the \lco\ coefficient has the same sign.
It seems that, in some sense, the CO content (not necessarily the \htwo\ content which depends also on metallicity as we have inferred it,
see Sect. \ref{sec:detailed}) is linked to $Z$, and can partially substitute the influence of \mstar. 
If we express \logoh\ in terms of \lco, as we have done in Eqn. (\ref{eqn:pca4dh2}), we obtain:

\begin{eqnarray}
o & = & (0.19\,\pm\,0.04)\,m - (0.19\,\pm\,0.03)\,s + (0.15\,\pm\,0.03)\,\ell \nonumber \\
\label{eqn:pca4dlco}
\end{eqnarray}

\noindent
where $o$, $m$, $s$ are as in Eqn. (\ref{eqn:pca4dh2}), and $\ell$ is the centered variable log(\lco). 
This expression is accurate to $\sim$0.11\,dex
assessed, as above for \mhtwo, by fitting a Gaussian to the residuals of the fit. 
In reality, this fit should not be interpreted rigorously, since the gas content, rather than CO luminosity,
is the quantity of interest.
The relation between \lco~ and the molecular gas content is almost entirely governed by metallicity \citep[e.g.,][]{Hunt2015,Accurso2017}; 
thus the strong dependence by \lco ~in Eqn. (\ref{eqn:pca4dlco}) is a reflection of the strong metallicity dependence of the conversion factor \aco. 
We will explore this notion more in detail in a future paper.

\begin{table}[!t] 
\caption{MAGMA sample means}
\label{tab:means}
\resizebox{\linewidth}{!}{
\begin{tabular}{lcccc}
\hline
\\ 
\multicolumn{1}{c}{Quantity} &
\multicolumn{1}{c}{Mean$^{\mathrm a}$} &
\multicolumn{1}{c}{Std. dev.} &
\multicolumn{1}{c}{Mean} &
\multicolumn{1}{c}{Mean} \\
&&& \multicolumn{1}{c}{(\mstar$\leq$} & \multicolumn{1}{c}{(\mstar$>$} \\
&&& \multicolumn{1}{c}{ \mbreak$^{\mathrm b}$)} & \multicolumn{1}{c}{ \mbreak)} \\
\multicolumn{1}{c}{(1)} &
\multicolumn{1}{c}{(2)} &
\multicolumn{1}{c}{(3)} &
\multicolumn{1}{c}{(4)} &
\multicolumn{1}{c}{(5)} \\
\\ 
\hline 
\\
log(\mstar/\msun)       & $9.684$     &  $0.75$           & $9.482$       & $10.565$  \\
log(SFR/\msunyr)        & $-0.229$    &  $0.66$           & $-0.375$      & $0.409$  \\
\logoh\                 & $8.580$     &  $0.22$           & $8.546$       & $8.728$  \\
log(\mhi/\msun)         & $9.279$     &  $0.70$           & $9.166$       & $9.774$ \\
log(\mhtwo/\msun)       & $8.720$     &  $0.69$           & $8.550$       & $9.463$ \\
log(\mgas/\msun)        & $9.438$     &  $0.64$           & $9.313$       & $9.985$ \\
log(\lco/               & $7.950$     &  $0.96$           & $7.728$       & $8.918$ \\
\quad\lcounits) \\
\\
\hline 
\end{tabular}
}
\begin{flushleft}
{\footnotesize
$^{\mathrm a}$~The number of galaxies considered in the mean for Cols. (2,3) is 392,
for Col. (4) 319, and for Col. (5) 73. \\
$^{\mathrm b}$~\mbreak\,=\,$2 \times 10^{10}$\,\msun\ (see also Fig. \ref{fig:mstar_coeffs}).
}
\\
\end{flushleft}
\end{table}

\begin{table}[!h] 
\caption{3D PCA results for MAGMA$^{\mathrm a}$}
\resizebox{\linewidth}{!}{
\label{tab:3dpca}
\begin{tabular}{lccccc}
\hline
\\ 
\multicolumn{1}{c}{Method} &
\multicolumn{1}{c}{PC3(1)} &
\multicolumn{1}{c}{PC3(2)} &
\multicolumn{1}{c}{PC3(3)} &
\multicolumn{1}{c}{PC3} &
\multicolumn{1}{c}{PC3} \\
& \multicolumn{1}{c}{log(\mstar/\msun)} &
\multicolumn{1}{c}{log(SFR/\msunyr)} &
& \multicolumn{1}{c}{std. dev.} &
\multicolumn{1}{c}{variance} \\
\multicolumn{1}{c}{(1)} &
\multicolumn{1}{c}{(2)} &
\multicolumn{1}{c}{(3)} &
\multicolumn{1}{c}{(4)} &
\multicolumn{1}{c}{(5)} &
\multicolumn{1}{c}{(6)} \\
\\ 
\hline 
\\
&&&\multicolumn{1}{c}{\logoh} \\
PCA  &  $0.346$          &  $-0.169$          & $-0.923$          & 0.127 & 0.015 \\
PPCA &  $0.29 \pm 0.02$  &  $-0.11 \pm 0.03$  & $-0.95 \pm 0.01$  & 0.14  & 0.02  \\ 
BSPCA & $0.34 \pm 0.02$  &  $-0.17 \pm 0.03$  & $-0.92 \pm 0.01$  & 0.13  & 0.02  \\ 
\\
&&&\multicolumn{1}{c}{log(\mhi/\msun)} \\
PCA  &   0.700            &  $-0.704$          & $-0.122$         &  0.264 & 0.047 \\
PPCA &   $0.69 \pm 0.03$  &  $-0.71 \pm 0.04$  & $-0.12 \pm 0.07$ &  0.29  & 0.06  \\ 
BSPCA &  $0.70 \pm 0.03$  &  $-0.70 \pm 0.05$  & $-0.13 \pm 0.08$ &  0.26  & 0.05  \\ 
\\
&&&\multicolumn{1}{c}{log(\mhtwo/\msun)} \\
PCA  &   0.680           &  $-0.730$          & $-0.061$        &  0.267 & 0.048  \\ 
PPCA &  $0.55 \pm 0.20$  &  $-0.68 \pm 0.19$  & $0.31 \pm 0.25$ &  0.29  & 0.06  \\ 
BSPCA & $0.61 \pm 0.18$  &  $-0.65 \pm 0.22$  & $0.33 \pm 0.23$ &  0.26  & 0.05  \\ 
\\
&&& \multicolumn{1}{c}{log(\mgas/\msun)} \\
PCA &   $0.742$         & $-0.613$         & $-0.272$         &  0.263 & 0.049  \\ 
PPCA &  $0.65 \pm 0.18$ & $-0.59 \pm 0.18$ & $-0.35 \pm 0.22$ &  0.29  & 0.06  \\ 
BSPCA & $0.70 \pm 0.09$ & $-0.60 \pm 0.16$ & $-0.30 \pm 0.18$ &  0.26  & 0.05  \\ 
\\
&&& \multicolumn{1}{c}{log(\lco/\lcounits)} \\
PCA   &   0.835           & $-0.356$          & $-0.420$         &  0.229 & 0.027   \\ 
PPCA  &  $0.82 \pm 0.04$  & $-0.42 \pm 0.11$ & $-0.36 \pm 0.08$  &  0.27  & 0.04  \\ 
BSPCA &  $0.83 \pm 0.01$  & $-0.36 \pm 0.08$ & $-0.42 \pm 0.06$  &  0.23  & 0.03  \\ 
\\
\hline 
\end{tabular}
}
\begin{flushleft}
{\footnotesize
$^{\mathrm a}$~As in Table \ref{tab:4dpca}, the relative signs of the PCs are arbitrary,
so that we have used the same conventions for all;
this has no bearing on the inversion of the equation of the PC with the least variance. 
Similarly to Table \ref{tab:4dpca}, 
Column (5) reports the standard deviation of PC3 around the hyperplane, and Col. (6) gives the proportion of its sample variance.
}
\\
\end{flushleft}
\end{table}

\subsection{3D PCA}
\label{sec:3dpca}

Section \ref{sec:4dpca} showed that the 4D parameter space can be approximated by a planar surface, with
$\ga$\,94\% of the variance contained in the first two eigenvectors, PC1$+$PC2.
Here we examine the 3D parameter space (in log space) by retaining \mstar\ and SFR as the two main observables,
and considering \logoh\ as one of the variables together with the four gas quantities
described above: \mhi, \mhtwo, \mgas, and \lco.
The aim of this exercise is to assess whether any of the gas parameters can be described only by \mstar\ and SFR,
and to investigate the implication of our 4D PCAs that metallicity \logoh\ can be adequately described by \mstar\
and SFR alone.

Using the same methodology as for the 4D 
PCA (``classic'' PCA without uncertainties, PPCA and BSPCA with uncertainties),
we have performed 3D PCAs on the MAGMA sample, and obtain the results reported in Table \ref{tab:3dpca}.
Like the 4D PCA, the 3D-PCA component with the least variance is dominated by the metallicity, \logoh\ 
(see column 4 in Table \ref{tab:3dpca}). 
Inverting the PC3 dominated by O/H as before for the 4D PCA, we find:
\begin{eqnarray}
o & = &(0.37\,\pm\,0.03)\,m - (0.18\,\pm\,0.03)\,s 
\label{eqn:pca3d}
\end{eqnarray}
The coefficients multiplying log(\mstar) and log(SFR) in Eqn. (\ref{eqn:pca3d}) are
the same to within the uncertainties as those from the 4D PCA given in Eqn. (\ref{eqn:pca4dh2}).
This expression describes \logoh\ for the MAGMA sample with an accuracy of $\sim 0.12$\,dex, again
obtained by fitting the residuals to a Gaussian.
The scatter of this expression is comparable to the scatter
obtained from the 4D PCA, leading to the conclusion that only \mstar\ and SFR are
necessary to describe metallicity.
The MAGMA coefficient for log(\mstar) of $0.37$ is the same as that found by \citet{Hunt2016a}.

\begin{figure*}[!t]
\includegraphics[width=0.95\textwidth]{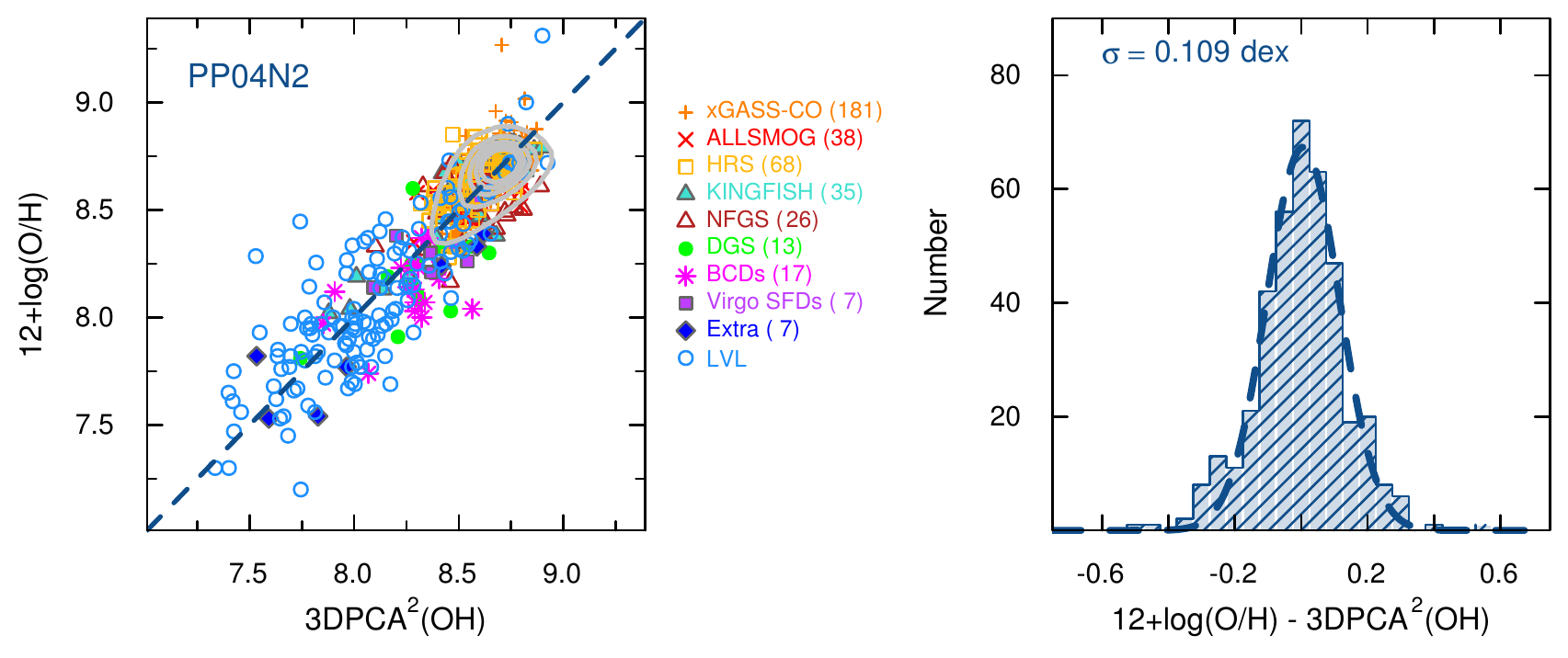}
\caption{Observed values of \logoh\ in the MAGMA sample compared to those predicted by the \pcatwo\ given in Eqn. (\ref{eqn:3dpcaoh2}). 
The parent sample of the individual MAGMA galaxies is given in the legend in the middle panel,
and the right panel shows the residuals and the Gaussian fit;
the 1$\sigma$ dispersion of the Gaussian is 0.074\,dex as discussed in the text. 
Also shown is the MAGMA \pcatwo\ applied to SDSS10, as before with 90\% of SDSS10 enclosed within the contours.
The median of the SDSS10 residuals relative to the \pcatwo\ is $0.002$\,dex, with a standard deviation of $\sim 0.08$\,dex,} 
showing that even with the MZR curvature clearly evident in SDSS10, the MAGMA \pcatwo\ does a good
job of reproducing the metallicities (see text for more details).
\label{fig:fpz}
\end{figure*}

There are two considerations here:
one is that a PCA, by definition, must pass through the multi-variable centroid of the dataset.
That is why here, in contrast to \citet{Hunt2012} and \citet{Hunt2016a}, we have defined the PCA results in terms of
centered variables.
This is important when applying a PCA determined with one sample to another sample; if the means of the
two samples are significantly different, then the PCA will not pass through the barycenter of the data for the second
sample, and will apparently not be a good fit.
{\it Thus a PCA must be applied using the centered variables associated with a particular data set. }
The second consideration is that despite the similarity in \mstar\ coefficients, 
the curvature of the MZR present in MAGMA (and SDSS10) is somewhat more pronounced than for the
sample analyzed by \citet{Hunt2016a}.
Fig. \ref{fig:mstar_coeffs} shows the coefficient of \mstar\
for the subsample with \mstar\,$\leq$\,\mbreak\ and \mstar\,$>$ \mbreak,
where \mbreak\ is the value of \mstar\ where one PCA ends and the other one starts.
The slopes for \mstar\ are systematically smaller for increasing \mbreak, because of the flattening of the MZR.
In the following section, we explore a remedy for this using an approach more appropriate for data showing non-linear relationships.

\subsubsection{3D PCA, a non-linear approach}

Several methods have been developed to assess mutual dependencies and dimensionality in a dataset that shows non-linear behavior.
In particular, curvature in a dataset can be first approximated by a piecewise linear approach
\citep[e.g.,][]{Hastie1989,Strange2015,Xianxi2017}. 
In the case of the curved MZR and its relation with SFR \citep[e.g.,][]{Mannucci2010,Cresci2018}, this is 
a fairly good approximation as we show below.
The fit to the MZR given by \citet{Andrews2013}, shown in Fig. \ref{fig:mzr}, consists of a mainly linear
portion toward low \mstar, connected smoothly to a roughly flat regime at high \mstar\
\citep[see also][]{Curti2019}.
Thus, as a simplified solution to the problem of MZR curvature, we approximate its behavior with two linear segments,
and perform a PCA separately on each.
Such a procedure is a specific example of a more complex piecewise linear approach, and we postpone a more
detailed analysis to a future paper.

The only ``free parameter'' in the piecewise linear PCA exercise is the break mass, \mbreak, namely the value of \mstar\ where we establish
the transition from one PCA to the other.
We have investigated \mbreak\ between $3\times10^9$\,\msun\ and  $3\times10^{10}$\,\msun\ (see Fig. \ref{fig:mstar_coeffs}) 
and find a ``sweet spot'' around \mbreak\,=\,$2\times 10^{10}$\,\msun, where the overall variance is minimized.
The best-fit piecewise 3D-PCA for MAGMA is as follows:
{\small
\begin{equation}
o\ = \ \begin{cases}
(0.43\,\pm\,0.03)\,m - (0.21\,\pm\,0.04)\,s & \text{if log(\mstar/\msun) $\leq$ 10.3} \\
(0.25\,\pm\,0.11)\,m - (0.11\,\pm\,0.05)\,s & \text{if log(\mstar/\msun) $>$ 10.3} \\
\end{cases}
\label{eqn:3dpcaoh2}
\end{equation}
}

\noindent
and the averages of the parameters in the two bins are given in Table \ref{tab:means}.

Figure \ref{fig:fpz} shows the piecewise 3D-PCA results [hereafter ``\pcatwo'']
where we compare the predictions of \logoh\ from Eqn. (\ref{eqn:3dpcaoh2}) and the means given above
to the observed values (vertical axis).
The parent samples of the individual MAGMA galaxies are given in the legend in the middle panel.
The standard deviation of the PC3 component is slightly smaller (0.11\,dex vs. $\sim$0.12\,dex) 
than that in the metallicity-dominated 4D PCA with, however, gas content taken into account.
The piecewise PC3 standard deviation of 0.11\,dex is also slightly smaller than the continuous 3D PCA result without gas, $\sim$0.12\,dex.
The Gaussian fit to the \pcatwo\ residuals is shown in the right panel of Fig. \ref{fig:fpz}. 
We expect that the degree to which the piecewise is better than the single PCA
depends on the number of galaxies more massive than $\sim 2\times10^{10}$\,\msun, i.e., the amplitude of the curvature in the MZR.

Also reported in Fig. \ref{fig:fpz} is the SDSS10 sample, to which we have applied the \pcatwo\ determined from MAGMA;
the grey contours enclose 90\% of the sample.
The mean (median) SDSS10 residuals are  $0.000$ ($0.002$) \,dex with a standard deviation of 0.08\,dex over 78579 galaxies.
Thus the MAGMA \pcatwo\ applied to SDSS10 represents the metallicities in that sample with accuracy 
comparable to the scatter found for the new formulation of the FMR for SDSS by \citet{Curti2019},
and with low systematics given the zero mean.
For LVL, the scatter is slightly worse: 
mean (median) LVL residuals are $0.000$ ($-0.02$)\,dex, with a standard deviation of $\sim$0.2\,dex over 135 galaxies.
Nevertheless, the small mean (median) residuals indicate that the LVL metallicities are also fairly well approximated
by the MAGMA \pcatwo\ even for the low masses in LVL, with a median log(\mstar/\msun) of 8.1, and 25\%
of the galaxies less massive than log(\mstar/\msun)\,=\,7.3.

Ultimately, 
comparison of the 4D and 3D PCAs shows 
that there is no need to include gas content, either \mgas, \mhi, or \mhtwo, in the description of $Z$;
it is statistically irrelevant, since a similar scatter is obtained without any gas coefficient. 
{\bf This is a clear confirmation that metallicity in field galaxies in the Local Universe
can be determined to $\la$0.1\,dex accuracy using only \mstar\ and SFR.
}
However, it is not a statement that metallicity is independent of gas content; 
on the contrary, 
in a companion paper,
we describe how gas content shapes the MZR through star-formation-driven outflows.
As we shall see below, the point is that gas content, like metallicity, can be described through \mstar\ and SFR dependencies.

\section{Comparison with previous work}
\label{sec:comparison}

Our results are in stark contrast with those of \citet{Bothwell2016b} who, as mentioned above, 
in a 4D PCA found that \htwo\ mass had a stronger link with metallicity than SFR.
\citet{Bothwell2016a} found a similar result based on a 3D PCA, namely that gas content drives
the relation between \mstar\ and metallicity, and that any tertiary dependence on SFR is merely 
a consequence of the Schmidt-Kennicutt relation between gas mass and SFR. 
In a similar vein,
\citet{Brown2018} through stacking
and \citet{Bothwell2013} 
found that \hi\ mass is strongly tied to $Z$, more than to SFR, similarly to later results for \mhtwo.
We conclude, instead, that metallicity is more tightly linked with stellar mass and SFR than
with either \mgas, \mhi, or even \mhtwo.
There are several possible reasons for this disagreement, and we explore them here,
with additional details furnished in Appendix \ref{app:cautionary}.

\subsection{Metallicity calibration and CO luminosity-to-molecular gas mass conversion}
\label{sec:bothwellcalibration}

We first examine how our results change if we use the same metallicity calibration
as \citet{Bothwell2016a,Bothwell2016b}.
This is potentially an important consideration because the KD02 O/H calibration
used by Bothwell et al. tends to give metallicities that are too high
\citep[e.g.,][]{Kewley2008}, relative to direct-\te\ estimates;
as shown in Fig. \ref{fig:mzr},
the PP04N2 is a better approximation of these \citep{Andrews2013,Hunt2016a,Curti2017}.
Together with using the KD02 calibration, we have also assessed
the effect of applying the \aco\ conversion factor used by Bothwell et al..
The exponential metallicity dependence proposed by \citet{Wolfire2010,Bolatto2013}
depends more steeply on metallicity than the power-law dependence we have used
above, as formulated by \citet{Hunt2015}.
Thus it is possible that the effects of metallicity are enhanced for the molecular gas mass \mhtwo\
with this approach.


We have thus applied these calibrations to the MAGMA sample, and performed a 4D PCA, as in \citet{Bothwell2016b}. 
The results of this exercise are reported in the lower portion of Table \ref{tab:4dpca}.
With the KD02 calibration and the exponential metallicity dependence of \aco,
we find that the PCA4 coefficients are slightly altered:
the \mstar\ and SFR coefficients are larger in amplitude and O/H coefficient is smaller.
The \htwo\ term is even smaller than with our original formulation, and zero to within the uncertainties.
In agreement with our original formulation,
we would have concluded that \htwo\ has negligible impact relative to SFR.
Thus, the different approaches for \aco\ and the metallicity calibration are probably not the cause
of the disagreement.

\subsection{Differences in sample sizes and properties}
\label{sec:bothwellsample}

Here we examine whether the larger MAGMA sample, its significant low-mass representation,
and different SFR relations can influence PCA results.
Our MAGMA sample of 392 galaxies is nominally twice as large as the sample studied by \citet{Bothwell2016a,Bothwell2016b}.
However, if we consider only the CO detections in their low-$z$ sample (141 galaxies), 
judging from Table 2 in \citet{Bothwell2016a}, our sample is almost three times larger.
Moreover, MAGMA contains a much larger fraction of low-mass galaxies, as it includes the
HeViCs dwarf galaxies \citep{Grossi2015}, the DGS \citep{Cormier2014}, the BCDs not yet published
by Hunt et al., and DDO\,53, Sextans\,A, Sextans\,B and WLM, the extremely metal-poor galaxies studied by \citet{Shi2015},
\citet{Shi2016} and \citet{Elmegreen2013}. 
The MAGMA mean log(\mstar/\msun)\,=\,9.7 is
$\sim$ 3 times lower than the mean log(\mstar/\msun)\,=\,10.2 of the 158 (including high-$z$) 
detections in the \citet{Bothwell2016a} sample;
while 24\% (94) of the MAGMA galaxies have \mstar$\leq 10^{9.3}$\,\msun,
this is true for only 7\% (11) of the \citet{Bothwell2016a} galaxies, and for $\sim$11\% (18) of those in \citet{Bothwell2016b}.

Nevertheless, the most important difference between the MAGMA sample and the Bothwell et al. sample(s) is
the inclusion of galaxies at high redshift in the latter.
As shown in Appendix \ref{app:cautionary},
the addition of these galaxies significantly increases the amplitude of the \mhtwo\ term in the 4D PCA,
and reduces that of SFR.
When the $z \sim 2$ galaxies are not included, the results of a 4D PCA on the Bothwell et al. sample are ambiguous, because
the metal content is found to increase with increasing SFR, similarly to the increase with \mstar.
However the statistical significance of this result is low, and the sample is ill conditioned because of the
behavior of SFR with \mstar\ in the sample.

\subsection{Methodology comparison and parameter uncertainties}
\label{sec:methods}

In Appendix \ref{app:injection}, we assess the consequences of introducing Gaussian noise
on an observing sample that is to be subject to a PCA.
After constructing several sets of mock samples based on well-defined input scaling relations,
we conclude that the accuracy with which the original relations can be retrieved depends
on the amount of noise injected.
It is fairly common to calculate uncertainties on fitted parameters by injecting noise in a sample 
and repeating the exercise several times \citep[e.g.,][]{Bothwell2016a,Bothwell2016b}.
However, our results show that this process skews the data because of the 
broader range in the parameter space, and relative importance of outliers in a PCA.

Another important consideration is the importance of the \mstar\ distribution of a sample
like the one considered here.
At a given level of noise injection $\sigma$, we found that the broader the range of \mstar, 
the more consistent with the original ``true'' scaling relations will be the results.

In some sense, as we show in Appendix \ref{app:injection}, observing samples such as MAGMA
already contain noise, and adding more will skew results, compromising reliability.
Ultimately, these are the reasons we chose to apply probabilistic PCA and boot-strap PCA with
sample replacement, rather than perturb the parameters of the sample by injecting noise.

\section{Summary and conclusions}
\label{sec:conclusions}

With the aim of investigating the role of gas on the mass-metallicity relation,
we have compiled a new `MAGMA' sample of 392 galaxies covering unprecedented ranges 
in parameter space,
spanning more than 5 orders of magnitude in \mstar, SFR, and \mgas,
and almost 2 orders of magnitude in metallicity. 
Basic galaxy parameters, \mstar\ and SFR, have been recalculated using
available data from IRAC, WISE, and GALEX archives, and
all O/H values have been converted to a common metallicity calibration, PP04N2.
All stellar masses and SFRs rely on a common \citet{Chabrier2003} IMF, 
and the combined sample has been carefully checked for potential
systematics among the sub-samples. 

Applying 4D and piecewise 3D PCAs to MAGMA confirms previous results that
O/H can be accurately ($\lesssim 0.1$\,dex) described only using \mstar\ and SFR.
However, our findings contradict earlier versions of PCA dimension reduction on smaller samples,
as we find that the O/H depends on SFR more strongly than on either \hi\ or \htwo.
Thus, even though a PCA shows mathematically that only a 2D plane is necessary to describe metallicity $Z$ or \mgas\ (or \mhtwo),
the dependence of $Z$ on gas content is not well constrained with a PCA.

In Sect. \ref{sec:4dpca}, a 4D PCA showed that the four parameters \mstar, SFR, \logoh, and \mhtwo\
or \mgas\ are related through a 2D planar relation,
with metallicity as the main primary dependent variable.
{\bf This implies that O/H depends primarily on \mstar\ and SFR, but also that
\mgas\ must depend primarily on these two variables because of the physical connection
between gas content and metallicity.}

The observational scaling relations here for O/H are applicable to isolated (field)
galaxies in the Local Universe, over a wide range of stellar masses, SFR, and metallicities \logoh\,$\ga$\,7.6.
They can be used as a local benchmark for cosmological simulations and to calibrate evolutionary
trends with redshift.
Future papers will consider relations among gas content, star-formation, and metal-loading efficiencies, as well as detailed
comparison with evolutionary models.

\section*{Acknowledgements}
We are grateful to the anonymous referee whose critical comments improved the paper.
The authors would also like to thank A. Baker, F. Belfiore, S. Bisogni, C. Cicone, M. Dessauges-Zavadsky, Y. Izotov, S. Kannappan, R. Maiolino, P. Oesch and D. Schaerer for helpful discussions.
We acknowledge funding from the INAF PRIN-SKA 2017 program 1.05.01.88.04.
We thank Yong Shi for passing us their results in digital form. 
We have benefited from the public available programming language \texttt{Python}, including the \texttt{numpy}, \texttt{matplotlib} and \texttt{scipy}  packages.
This research made extensive use of \texttt{ASTROPY}, a community-developed core \texttt{Python} package for Astronomy \citep{Astropy2013}, and \texttt{glueviz}, a \texttt{Python} library for multidimensional data exploration \citep{Beaumont2015}.
This research has also made use of data from the HRS project; 
HRS is a \hers\ 
Key Programme utilizing Guaranteed Time from the SPIRE instrument team, ESAC scientists and a mission scientist.
The HRS data was accessed through the Herschel Database in Marseille (HeDaM - http://hedam.lam.fr) operated by CeSAM and hosted by the Laboratoire d'Astrophysique de Marseille.

\bibliographystyle{aa}
\bibliography{biblio} 

\appendix
\section{Overall comparison of galaxy parameters in MAGMA}
\label{app:comparison}

\begin{figure}[!h]
	\centering
	\includegraphics[width=0.92\linewidth]{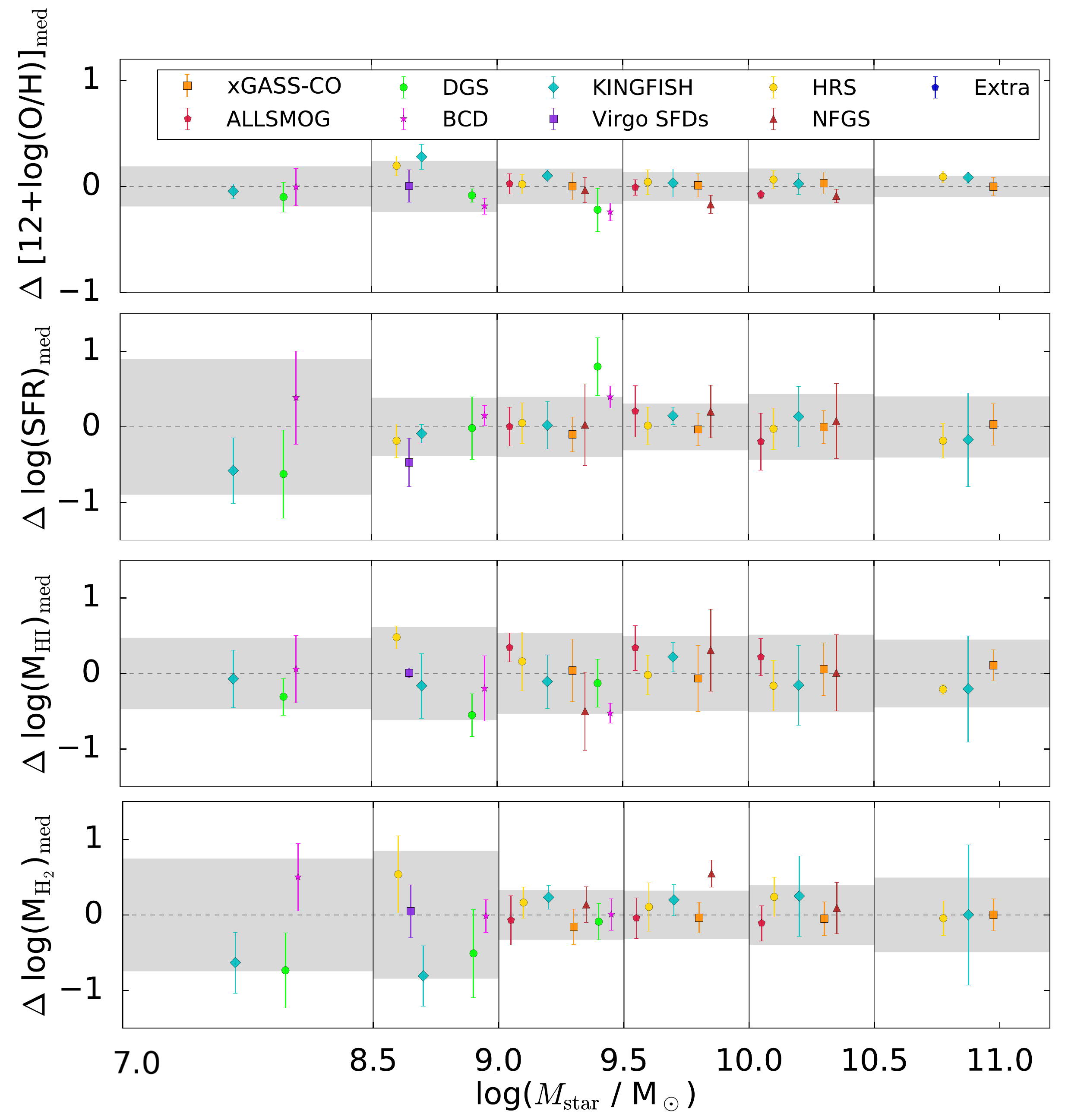}
	
	\caption{Comparison of SFR, \logoh, 
		\mhtwo, and \mhi\ in segregated \mstar\ bins for the MAGMA sample.
		The grey regions correspond to the standard deviations ($\pm 1\sigma$),
		and the horizontal dashed line to the zero difference (by definition) relative to the sample
		median.
		The 9 different parent samples comprising MAGMA are shown by different symbols
		given in the legend in the upper right corner.
		The points and their error bars correspond to the median of the difference,
		relative to the median parameter of the MAGMA sample as a whole.
	}
	\label{fig:comparison}
\end{figure}

Figure \ref{fig:comparison} shows the median differences of each individual
parent samples in MAGMA relative to the sample as a whole as a function of discrete
\mstar\ bin; the error bars are the standard deviations of the estimate.
The grey regions give the standard deviations of the entire MAGMA sample, and the
horizontal dashed lines the median difference, zero by definition.
If there were any systematic differences as a function of parent sample and/or \mstar,
they should stand out in Fig. \ref{fig:comparison}.
However, it is clear from the figure that there are no systematic differences in
any of the parameters shown. 
The only exception could be the \logoh\ (bottom) panel, where the BCD, DGS, Virgo SFDs
have O/H lower than would be expected; there is a similar corresponding excess, although within the spreads,
in SFR. 
This is a real, physical difference, among the samples, due to the anti-correlation between SFR and O/H, the FMR
\citep[see, e.g.,][]{Mannucci2010}.
These dwarf samples at low \mstar\ are slightly starburst biased because of our requirement for CO detections.
Thus, with this exception, we conclude that there are no apparent systematic differences among the parent samples in MAGMA.

\section{Details of comparison with previous work: a cautionary tale}
\label{app:cautionary}

\begin{figure}[!h]
\includegraphics[width=\linewidth]{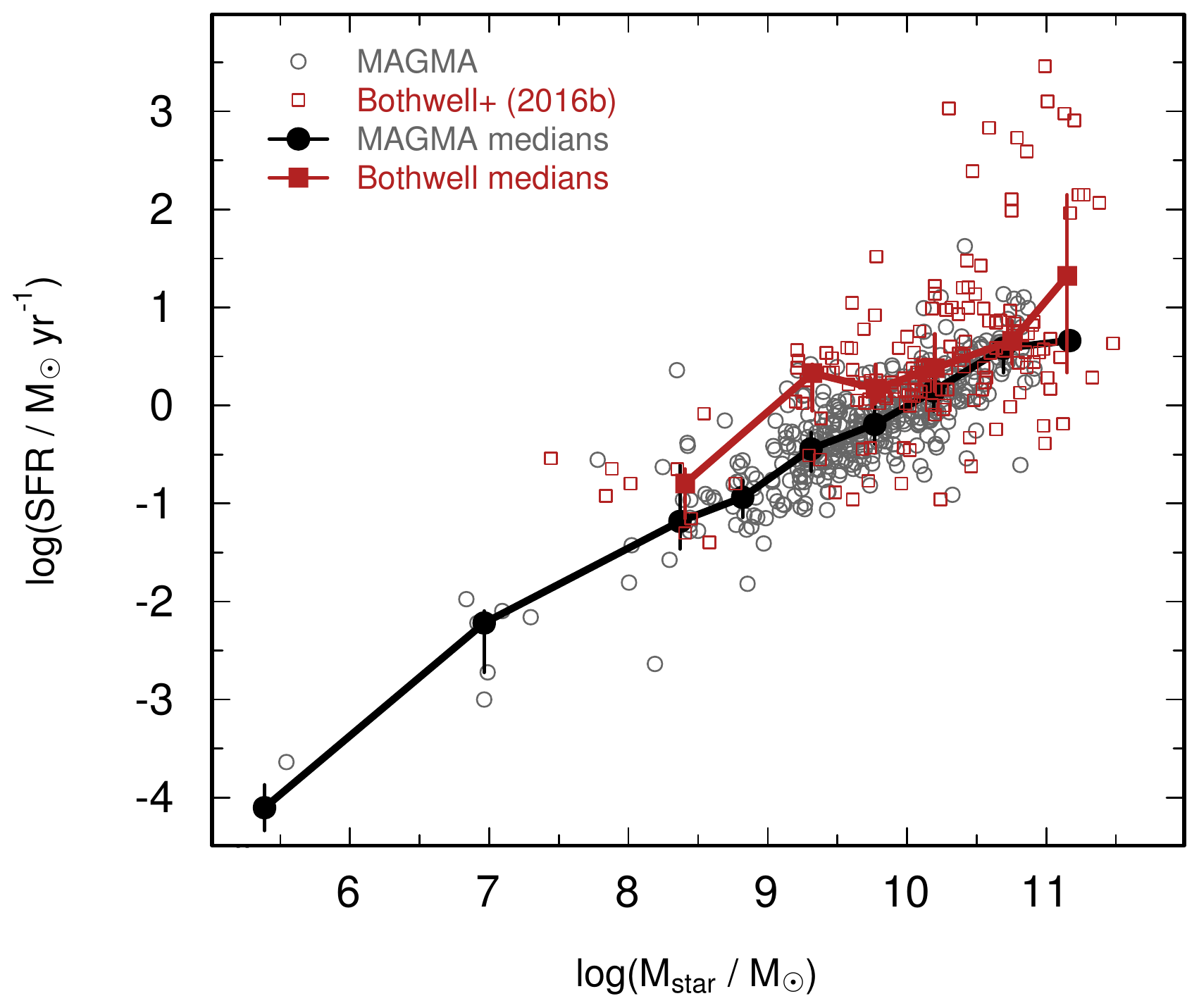}
\caption{Log(SFR) vs. Log(\mstar) for the MAGMA sample as in Fig. \ref{fig:scaling1},
but here also plotted are the galaxies from \citet[][see their Table 2]{Bothwell2016a},
together with the 8 galaxies from \citet{Hunt2015} added but with reduced \mstar\
as described in the text.
Medians for both samples are also shown, but 
only medians with $\geq\,3$ points in the respective \mstar\ bins are plotted here.
}
\label{fig:bothwellsfr}
\end{figure}

\begin{table*}[!t] 
\caption{4D PCA results for the \citet{Bothwell2016b} sample$^{\mathrm a}$}
\resizebox{\linewidth}{!}{
\label{tab:bothwellpca}
\begin{tabular}{lcccccccc}
\hline 
\\
\multicolumn{1}{c}{Method} &
\multicolumn{1}{c}{PC4(1)} &
\multicolumn{1}{c}{PC4(2)} &
\multicolumn{1}{c}{PC4(3)} &
\multicolumn{1}{c}{PC4(4)} &
\multicolumn{1}{c}{PC4} &
\multicolumn{1}{c}{PC4} &
\multicolumn{1}{c}{PC3} &
\multicolumn{1}{c}{PC1$+$PC2} \\
& \multicolumn{1}{c}{\logoh} &
\multicolumn{1}{c}{log} &
\multicolumn{1}{c}{log} &
\multicolumn{1}{c}{log} &
\multicolumn{1}{c}{std. dev.} &
\multicolumn{3}{c}{proportion of variance} \\
& \multicolumn{1}{c}{(KD02)} &
\multicolumn{1}{c}{(\mstar/\msun)} &
\multicolumn{1}{c}{(SFR/\msunyr)} & 
\multicolumn{1}{c}{(\mhtwo/\msun)$^{\mathrm b}$} \\
\\
\hline 
\multicolumn{9}{c}{{\it Including the 17 high-$z$ galaxies and 8 BCDs (166)}} \\
\hline 
\\ 
PCA   &  $0.872$           & $-0.408$            &  $-0.125$            & $0.242$            &  0.169 &  0.014 & 0.042 & 0.943\\
PPCA  &  $0.905 \pm 0.016$ & $-0.381 \pm 0.018$  &  $-0.069 \pm 0.038$  & $0.165 \pm 0.044$  &  0.182  & 0.017  \\ 
PPCA  &  $0.903 \pm 0.017$ & $-0.383 \pm 0.017$  &  $-0.071 \pm 0.042$  & $0.167 \pm 0.047$  &  0.182  & 0.017  \\ 
BSPCA &  $0.871 \pm 0.028$ & $-0.404 \pm 0.027$  &  $-0.125 \pm 0.042$  & $0.239 \pm 0.054$  &  0.167  & 0.014  \\ 
BSPCA &  $0.865 \pm 0.033$ & $-0.405 \pm 0.025$  &  $-0.131 \pm 0.048$  & $0.249 \pm 0.062$  &  0.168  & 0.014  \\ 
\\
\hline 
\multicolumn{9}{c}{{\it With 8 BCDs but without the 17 high-$z$ galaxies (149)}} \\
\hline 
\\
PCA   &  $0.891$           & $-0.378$            &  $-0.173$            & $0.186$            &  0.163 &  0.023 & 0.074 & 0.90\\
PPCA  &  $0.908 \pm 0.015$ & $-0.355 \pm 0.020$  &  $-0.155 \pm 0.038$  & $0.145 \pm 0.038$  &  0.173  & 0.027  \\ 
PPCA  &  $0.909 \pm 0.015$ & $-0.357 \pm 0.019$  &  $-0.150 \pm 0.039$  & $0.142 \pm 0.042$  &  0.173  & 0.027  \\ 
BSPCA &  $0.888 \pm 0.033$ & $-0.374 \pm 0.039$  &  $-0.171 \pm 0.046$  & $0.184 \pm 0.069$  &  0.159  & 0.023  \\ 
BSPCA &  $0.886 \pm 0.032$ & $-0.378 \pm 0.039$  &  $-0.169 \pm 0.042$  & $0.188 \pm 0.066$  &  0.160  & 0.023  \\ 
\\
\hline 
\end{tabular}
}
\begin{flushleft}
{\footnotesize
$^{\mathrm a}$~In PCA, the relative signs of the PCs are arbitrary, so that
we have used the same conventions for all; 
this has no bearing on the inversion of the equation of the PC with the least variance. \\
$^{\mathrm b}$~Here \mhtwo\ is calculated from \aco\ according to the exponential formulation of \citet[][]{Wolfire2010,Bolatto2013}.
}
\\
\end{flushleft}
\end{table*}

There are two salient differences between the MAGMA sample and the sample from \citet{Bothwell2016b}. 
One is the significant low-mass coverage of the MAGMA sample, and the other is the SFRs.
Both these are illustrated in
Fig. \ref{fig:bothwellsfr} where the MS for the two samples are plotted;
the median points show only those \mstar\ bins with $\geq\,3$ data points. 
The \citet{Bothwell2016b} sample is the same as that from \citet{Bothwell2016a}, but with 8 BCDs
from \citet{Hunt2015}; 
\citet{Bothwell2016b} altered the \mstar\ values, and 
to best approximate their sample, for the BCDs from \citet{Hunt2015} we have arbitrarily lowered
the \mstar\ values by a factor of three\footnote{\citet{Bothwell2016b} has used a different
technique to estimate \mstar, and find roughly a factor of 3 times lower values, but there is no tabulation
of their modified values.}.
Even with the addition of the BCDs from \citet{Hunt2015},
the mass distribution is significantly more extended for MAGMA.
The low-mass bin in the Bothwell et al. sample of Fig. \ref{fig:bothwellsfr} are the galaxies from \citet{Hunt2015} that are not
present in the sample used by \citet{Bothwell2016a}, and the highest-mass
bin is dominated by the high-$z$ sub-millimeter galaxies (SMGs) and MS galaxies at $z \sim 2$.
MAGMA shows slight downward inflection associated with quiescent or passive high-mass galaxies
(see also Fig. \ref{fig:scaling1}),
while the \citet{Bothwell2016b} sample shows a steep upturn, because of the inclusion of high-$z$ massive galaxies
with consequently higher SFR.

There are also differences in the SFRs: while MAGMA adopts COLDGASS galaxies with
parameters from \citet{Saintonge2017},
\citet{Bothwell2016a,Bothwell2016b} use the COLDGASS parameters reported by \citet{Saintonge2011a,Saintonge2011b}.
However, a subsequent study \citep{Huang2014} showed that the SFRs in those papers are overestimated by $\sim 0.2$\,dex because of 
aperture effects from the CO single-dish IRAM beam.
Roughly half (115 galaxies) of the Bothwell et al. sample is from COLDGASS which means that these values are also discrepant with respect to MAGMA. 
This can be seen in Fig. \ref{fig:bothwellsfr} where, at a given \mstar, the Bothwell et al. sample tends to have higher SFRs than MAGMA.

Possibly the most important difference in this context relative to MAGMA is the inclusion by Bothwell et al. of the 
17 high-$z$ galaxies (9 main-sequence galaxies and 8 SMGs).
These galaxies are at $z\sim 2$, and have significantly higher SFRs than the local
galaxies of similar stellar mass, because of the increasing normalization of the star-formation main sequence
with redshift \citep[e.g.,][]{Speagle2014}.
The minimum Log(\mstar/\msun) value in the high-$z$ sample is 9.78, and the mean SFR $\sim 274$\,\msunyr;
for the low-$z$ sample over the same mass range, the mean SFR $\sim 2.4$\,\msunyr,
more than 100 times smaller.
A similar difference applies to the ratios of \mhtwo\ in the two samples, where the mean \mhtwo\ in the high-$z$ sample
is $\sim 85$ times higher than in the low-$z$ galaxies over the same range in \mstar. 
To assess the impact of these galaxies on the results by \citet{Bothwell2016a,Bothwell2016b},
we have performed 4D PCAs on the \citet{Bothwell2016b} sample, 
with and without the 17 high-$z$ galaxies, using only the galaxies with CO detections in their Table 2 \citep[and][]{Hunt2015}. 
Results are reported in Table \ref{tab:bothwellpca}.

The upper part of Table \ref{tab:bothwellpca} shows the 4D PCA for \htwo\ for the \citet{Bothwell2016b}
sample including the 17 high-$z$ MS galaxies and SMGs and the 8 BCDs from \citet{Hunt2015}. 
The results
are in fairly good agreement\footnote{This sample is not quite the same as that analyzed
by \citet{Bothwell2016b} because the 8 galaxies from \citet{Hunt2015}
are included with an arbitrary factor of 3 lower \mstar, since \citet{Bothwell2016b} recalculated
their \mstar\ values but did not tabulate the new \mstar\ values. 
Moreover the numbers of galaxies do not apparently match;
here we are analyzing only the 158$+$8 CO detections given in Table 2 of \citet{Bothwell2016a} with 8 BCDs
\citep[with \mstar/3 from][]{Hunt2015}.}
with those of \citet{Bothwell2016b}.
The dependence of \logoh\ on \mhtwo\ is larger than that on SFR;
our probabilistic PCA estimates of the uncertainties show, however, that the 
coefficient for the \mhtwo\ dependence is determined with $< 3\sigma$ significance.

The lower part of Table \ref{tab:bothwellpca} shows instead the 4D PCA result for the
low-$z$ \citet{Bothwell2016a} sample, without the 17 high-$z$ galaxies, but with the 8 BCDs from \citet{Hunt2015}.
The \mstar\ dependence is significantly reduced, the SFR dependence is increased, and the \mhtwo\
dependence is consistent with 0.0 to within the errors.
Interestingly, the SFR and \mstar\ coefficients have the same sign, implying that increasing both 
SFR and \mstar\ will increase $Z$; this is contrary to the ``normal'' \pcaone\ behavior in which
at a given \mstar, increasing SFR tends to reduce $Z$.
The comparison in Table \ref{tab:bothwellpca}
of the two versions of the Bothwell et al. sample, including or
omitting the high-$z$ galaxies, shows that the PCA significantly changes, and that the most probable
driver of the lack of metallicity SFR dependence relative to \mhtwo\ is caused by the 
inclusion of $z \sim 2$ galaxies which, at a given \mstar, have a significantly higher molecular gas content and
SFR than galaxies in the Local Universe.

\section{Assessment of Monte Carlo error injection on PCA fits}
\label{app:injection}

To explore the effect of injecting Gaussian noise on a dataset subject to a PCA,
we generated several ``mock'' samples of the \pcaone\ for galaxies at $z \sim 0$.
To do this, we first distributed numbers of galaxies in \mstar\ bins with redshifts ranging from $z\,=\,0.0$
to $z\,=\,0.02$,
according to either a constant \mstar\ distribution or one that resembles the GSMF given by \citet{Baldry2012}.
Within each mass and (small) redshift bin, we selected randomly \mstar\ in order to more or less reproduce
the assumed distribution.
Then we derived a MS of star formation by imposing \citet{Speagle2014} at $z\,=\,0$, or equivalently adopting the
relation given by \citet{Hunt2019} for the KINGFISH sample.
We added a small (realistic) fraction of starbursts using the formulation of \citet{Sargent2012};
this approach separates galaxy populations according to sSFR, and approximates the distribution with two
Gaussians.
\citet{Sargent2012} further assume that the starburst fraction is independent of mass and redshift,
which may or may not be correct \citep[see e.g.,][]{Bisigello2018}. 
Finally, we related \logoh\ to \mstar\ and SFR via the \pcaone\ (FPZ) reported by \citet{Hunt2016a}.
This means that the basic input parameter is \mstar, which defines SFR through the MS with the addition of a small
fraction of starbursts; then \logoh\ is calculated based on \mstar\ and SFR.
We adopted a total mock sample size for both \mstar\ distributions of $\sim$12\,000 simulated galaxies.

\begin{figure}[!h]
\includegraphics[width=\linewidth]{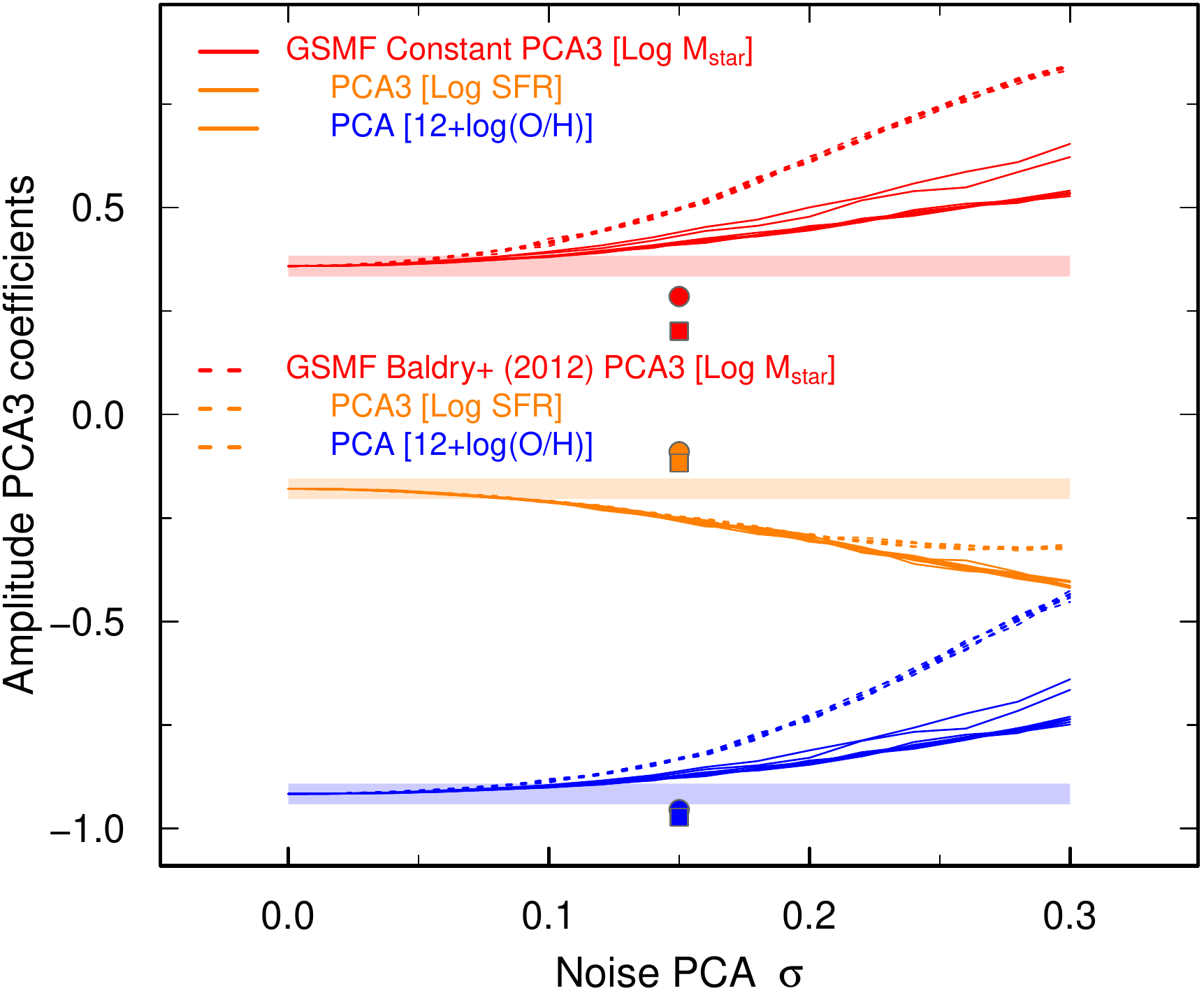}
\caption{3D PCA coefficients for Log(\mstar) (in red), Log(SFR) (in orange), and \logoh\ (in blue) plotted against the injected
noise level $\sigma$; different line types correspond to the two different \mstar\
distributions as described in the text and illustrated in the legend.
The different curves for the constant \mstar\ case correspond to different lower-mass limits.
The true input \pcatwo\ is shown by the transparent lines, with thickness of $\pm$0.025\,dex.
The data points (including error bars which are smaller than the symbols) report the PCA coefficients from 
the initial $\sigma\,=\,0.15$\,dex mock sample, but which has been in turn perturbed as described in
the text.
The aim of this subsequent exercise is to simulate a Monte Carlo perturbation of an observed sample.
}
\label{fig:pcanoise}
\end{figure}

This construction of the mock samples may seem arbitrary, but in truth the details are not important; 
we only want to compare what we get out with what we put in, in the case of varying levels of noise injection. 
To this end, we took this initial ``noiseless'' sample and introduced varying degrees of Gaussian uncertainty, $\sigma$,
to SFR and \logoh; for simplicity, we used the same value of $\sigma$ for both SFR and \logoh.
We repeated this procedure several times, and applied a PCA to each of the noise-injected samples.
The results are shown in Fig. \ref{fig:pcanoise},
where the 3D PCA coefficients are plotted against the injected noise level $\sigma$. 
The left-hand part of the curves for $\sigma\,=\,0$ are the input values of the \pcaone\ by which the mock sample was
generated.
There are seven separate curves in Fig. \ref{fig:pcanoise} for each of the three PCs, corresponding to seven
different realizations of the noise injection for the mock samples;
the closeness of the curves evident in the figure indicates that the statistical results are quite stable.
The discrepant curves for the constant \mstar\ case correspond to different lower-mass limits.

The idea here is to simulate an observed sample such as MAGMA, and assess the accuracy of the resulting PCA,
compared to the input ``true'' values.
The implicit assumption is that MAGMA, or similar samples, are governed by an underlying \pcaone\ or 2D plane,
but suffer from uncertainty in the measurement of the observables.
We cannot know whether or not this is true; we can only ascertain how far the observed data set could differ
from the underlying relation if it were present. 

Also shown in Fig. \ref{fig:pcanoise} are six data points corresponding to a PCA on the $\sigma\,=\,0.15$
mock sample, but to which an additional perturbation has been applied.
We have chosen $\sigma\,=\,0.15$\,dex as a starting point, in order to best approximate the behavior of the MAGMA sample which shows
a dispersion in \logoh\ of roughly this amplitude around the best-fit \pcaone. 
While the original mock samples are intended to reproduce observed samples, this additional injection of Gaussian
noise is aimed at simulating a Monte Carlo perturbation of an observed sample.
Thus, we injected Gaussian noise of various amplitudes on the variables of our mock sample:
Log(\mstar) was varied by 0.3\,dex;
SFR was varied by 30\% [corresponding to $\sim 0.2$\,dex on Log(SFR)];
\logoh\ was varied by 0.1\,dex.
This is a similar technique to that described by \citet{Bothwell2016a,Bothwell2016b}, and in principle helps to
establish uncertainties in the final PCA results.
However, Fig. \ref{fig:pcanoise} shows clearly that the injection of additional noise on the mock sample carries
the PCA results even further from the input true \pcaone.
The amplitude of the noise injected $\sigma$ already masks the \pcaone\ that was the basis for the mock samples,
but the additional perturbation exacerbates even more the capacity of the PCA to diagnose the underlying relation.

\begin{figure}[!h]
\includegraphics[width=\linewidth]{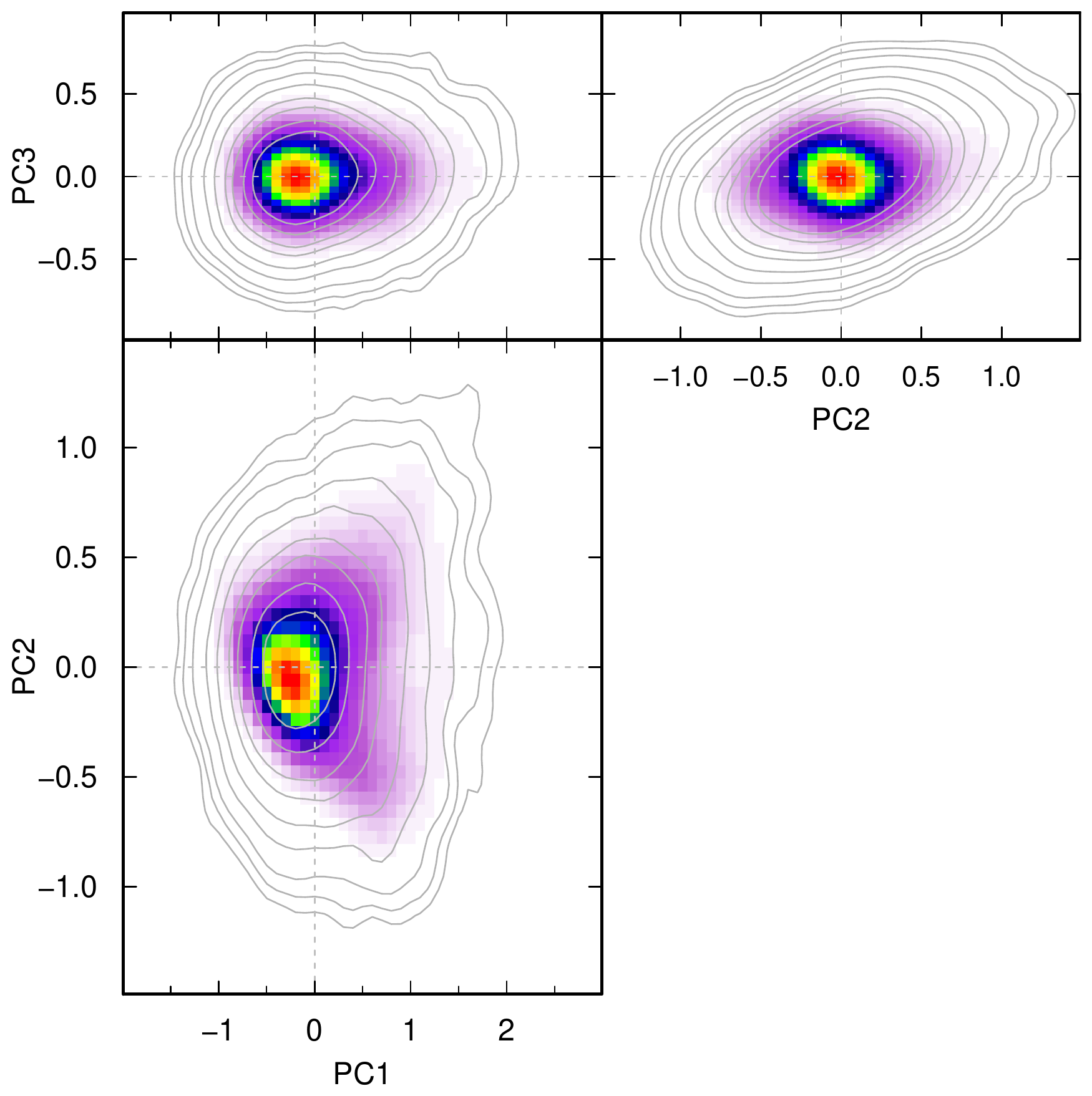}
\caption{3D PC components under various projections.
The underlying color maps show the density distribution of data points with the PCA calculated
from the original mock sample with $\sigma\,=\,0.15$\,dex,
and the contours of the density distribution of the perturbed sample but with the PCA loadings
of the original one.
The change of orientation of the PC decomposition introduced by the perturbation is evident, especially in the upper right panel. 
}
\label{fig:pcaperturbcomp}
\end{figure}

Fig. \ref{fig:pcaperturbcomp} illustrates in another way the process of this subsequent perturbation on the
mock sample; the orthogonal projections of the PCs are shown in three different panels.
The color maps show the original mock sample with $\sigma\,=\,0.15$ injected Gaussian noise,
while the contours show the same data as the individual points \citep[with the GSMF from][]{Baldry2012}, but here
with the PCs calculated according to the loadings of the original mock sample.
This is done to highlight the change in orientation relative to the original sample, thus illustrating that
the injection of additional noise  alters the orientation of the components. 
This can be seen in particular in the upper right panel which compares PC3, dominated by O/H, to PC2, which is dominated by \mstar,
and to a lesser extent SFR; the contours are oriented at a different angle relative to the underlying color map.

We conclude that:
\begin{itemize}
\item
The injection of Gaussian noise in a noiseless sample changes the PCA characteristics, because
of the resulting change of the orientation of the derived PCs (see the curves in Fig. \ref{fig:pcanoise}).
Further injection of Gaussian noise moves the PCA even further from the input relation,
as shown by the individual points in  Fig. \ref{fig:pcanoise}.
Even though the introduction of noise does not change the means of the parameters,
it skews the orientation because PCAs consider the entire distribution of data, including outliers.
\item
The distribution of \mstar\ in a sample also impacts the results of a PCA (see varying curves in Fig. \ref{fig:pcanoise}).
This is because a PCA calculates the orthogonal distance from an orientation, and requires
a broad distribution in parameters in order to better take into account eventual outliers.
\end{itemize}

Ultimately, because of the mathematical nature of the PCA, the addition of noise to a sample for which
a PCA is to be performed is deleterious for the reliability of the results.
For this reason, we have preferred here to use resampling techniques, rather than alter the noise characteristics
of the sample.

\end{document}